\documentclass[aps,prd,preprint,superscriptaddress,nofootinbib,preprintnumbers]{revtex4-1}

\usepackage{amsfonts,amsmath,amsthm,amssymb}
\usepackage{mathrsfs}
\usepackage{bm}
\usepackage{color}
\usepackage{axodraw4j}
\usepackage{float}
\usepackage{epsfig}
\newcommand{\PRE}[1]{{#1}}   % Use if preprint style

\newcommand{\comment}[1]{}
\newcommand{\ke}{\rangle}
\newcommand{\br}{\langle}

\newcommand{\p}{\partial}
\newcommand{\ba}{\begin{eqnarray}}
\newcommand{\ea}{\end{eqnarray}}
\newcommand{\be}{\begin{equation}}
\newcommand{\ee}{\end{equation}}
\newcommand{\eg}{\textit{e.g.} }
\newcommand{\ie}{\textit{i.e.} }
\newcommand{\cf}{\textit{cf.} }

\allowdisplaybreaks
\newcommand{\nocontentsline}[3]{}
\newcommand{\tocless}[2]{\bgroup\let\addcontentsline=\nocontentsline#1{#2}\egroup}

\newcommand{\req}[1]{(\ref{#1})}
\def\vev#1{\langle #1 \rangle}

\def\si{\sigma}

\def\fc#1#2{\frac{#1}{#2}}
\def\h{\frac{1}{2}}

\newcommand{\nwc}{\newcommand}

\def\ap{\alpha'}

\nwc{\bea} {\begin{eqnarray}}
\nwc{\eea} {\end{eqnarray}}
\nwc{\nnn} {\nonumber \\[5mm] }

\nwc{\bda} {\bdm\ba{lcl}}
\nwc{\eda} {\ea\edm}

\def\vev#1{\langle #1 \rangle}

\nwc{\ds}  {\displaystyle}
\nwc{\ra}{\rightarrow}
\nwc{\lra}{\longrightarrow}
\def\lf{\left}\def\ri{\right}

\def\Pc{{\cal P}}\def\Oc{{\cal O}}

\def\IC{{\bf C}}\def\IR{ {\bf R}}
\def\Mc{{\cal M}}
\def\Gc{{\cal G}}
\def\si{{\sigma}}\def\al{{\alpha}}\def\de{{\delta}}

\begin{document}

\preprint{
\hfil
\begin{minipage}[t]{4in}
\begin{flushright}
\vspace*{.3in}
MPP--2013--18\\
%CERN-PH-TH/2012-xxx\\
\end{flushright}
\end{minipage}
}

\vspace*{-5cm}
\title{Superstring Amplitudes as a Mellin Transform of Supergravity
\PRE{\vspace*{0.3in}} }

\author{Stephan Stieberger}
\affiliation{Max--Planck--Institut f\"ur Physik\\
 Werner--Heisenberg--Institut,
80805 M\"unchen, Germany
\PRE{\vspace*{.1in}}
}
\author{Tomasz R. Taylor}
\affiliation{Department of Physics\\
  Northeastern University, Boston, MA 02115, USA \PRE{\vspace*{.1in}}
}

%\date{February 2013}

\PRE{\vspace*{3.5in}}

\begin{abstract}
  \noindent
At  the tree level, the maximally helicity violating amplitudes of $N$ gauge bosons in open superstring theory
and of $N$ gravitons in supergravity  are known to have simple representations in terms of tree graphs.
{}For superstrings, the graphs encode integral representations of certain generalized Gaussian hypergeometric functions
of kinematic invariants while for supergravity, they  represent specific   kinematic
expressions constructed from
spinor--helicity variables. We establish a superstring/supergravity correspondence for
this class of amplitudes,
by constructing a mapping between the positions of gauge boson vertices at the disk boundary and the
helicity spinors associated to gravitons.
After replacing vertex positions by a larger set of
$\tfrac{N(N-3)}{2}$ coordinates,
the superstring amplitudes become (multiple) Mellin transforms of supergravity amplitudes,
from the projective space into the dual Mellin space of $\tfrac{N(N-3)}{2}$ kinematic invariants.
Similarly, inverse Mellin transforms transmute open superstrings into supergravity.
We elaborate on the  properties of multiple Mellin  and inverse Mellin transforms in the framework of superstring/supergravity correspondence.

\end{abstract}

\maketitle

\tableofcontents
\break

\section{Introduction}

Superstring scattering amplitudes are often considered far more involved than scatter{\nolinebreak}ing amplitudes in quantum field theory. There are many efficient perturbative techniques avail{\nolinebreak}able in quantum field theory, based on Feynman diagrams, recursion relations {\em etc.}, and even some non--perturbative aspects of scattering amplitudes
 can be studied by using the AdS/CFT correspondence. Furthermore,  a unified mathematical framework encompassing the complete perturbative S--matrix (all loops, arbitrary number of external particles)  in terms of the Grassmannian description, has been proposed for maximally supersymmetric gauge theories \cite{ArkaniHamed:2012nw}.

In string theory, although there has been some steady progress over the last thirty years, two--dimensional world--sheet conformal
field theory (CFT) still remains as the basic tool for computing scattering amplitudes. A rare newcomer to this research field has to digest several textbook chapters before even trying to reproduce the four--tachyon amplitude  written by Veneziano in 1968 \cite{Veneziano:1968yb}. In order to compute the lowest order, semi--classical scattering amplitude  of four gauge bosons in open superstring theory, one considers a disk world--sheet with four vertex operators inserted at the boundary. There is an integral to be performed, over the position of one of vertex operators, the other three being fixed by  M{\"{o}}bius transformations. This integral yields a ``special'' function of kinematic invariants, the Euler beta function which summarizes virtual exchanges of  gauge bosons and of the infinite tower of their string (Regge) excitations in all kinematic channels, but avoiding double--counting and  implementing the world--sheet duality of the old ``dual resonance model.''

More {\em multiple\/} integrals appear in the scattering amplitudes involving larger numbers of external gauge bosons (gluons) \cite{Oprisa:2005wu,Stieberger:2006te}.  In $N$--gluon amplitudes, kinematics are specified by $N(N-3)\over 2$ Lorentz invariants, instead of just two  Mandelstam's variables (for $N=4$), and there are $N{-}3$ vertex positions to be integrated over the boundary. As a result, one obtains many generalized {\em hypergeometric\/} functions of {\em many} kinematic variables, instead of a single beta function of two variables. In spite of such complications, some significant progress has been accomplished over the last few years. Most notably, $N$--gluon
superstring disk amplitudes have been expressed in terms of tree--level Yang--Mills amplitudes and $(N{-}3)!$ hypergeometric functions \cite{Mafra:2011nv,Mafra:2011nw}. More recently, a particularly simple formula has been derived for the maximally helicity violating (MHV) amplitudes \cite{Stieberger:2012rq}, with the functions represented by tree graphs.

This paper begins with a simple observation of a similarity between semi--classical MHV amplitudes
describing $N$ gauge bosons in open superstring theory and $N$--graviton MHV amplitudes in quantum field theory of supergravity. This similarity becomes most apparent when interpreting the amplitudes in terms of tree graphs. {}For superstrings, the vertices correspond to vertex positions at the disk boundary while for supergravity, they label the gravitons. Nevertheless, as shown in Section 2, the edge factors are identical, and both amplitudes can be expressed as certain minors of the same (generalized) Laplacian matrix. The only difference between supergravity and superstring is that in the latter case, there remain non--trivial integrations to be performed over the vertices.

In Section 3, we focus on the vertex integrations of superstring amplitudes. We uplift the vertex positions to a larger, ${N(N-3)\over 2}\,$--dimensional space, parameterized by M\"{o}bius invariant cross--ratios, in one-to-one correspondence with the kinematic invariants. The vertex integrations are lifted to the embedding projective space, to a surface localized by insertions of appropriate delta function constraints. We show that these integrals amount to a
multi--dimensional Mellin transform of the supergravity amplitude, from Mellin position space of M\"{o}bius invariant cross--ratios to the dual space of kinematic variables.

In Section 4, we elaborate general aspects of multiple Mellin transforms in the string framework.
We consider superstring amplitudes as Mellin amplitudes.
An integral transform (multiple inverse Mellin transform) of the latter yields simple expressions in terms of products of delta--functions localizing on the two--dimensional world--sheet.
Momentum dependence is reinstalled by applying  Mellin transforms.

In Section 5, we point to future directions and discuss some broader implications of our results.

In Appendix A, we discuss  polynomial reduction of a certain class of rational functions, which can be related to $(N-2)^{(N-4)}$ labelled trees on $N-2$ vertices (Cayley graphs). The latter
serve as a basis for writing both the $N$--point graviton supergravity amplitude and the $N$--point superstring gluon amplitude.
We prove that partial fraction decomposition reduces this set of rational functions
to a basis of $(N-3)!$ elements, which are graphically related to Hamilton graphs.
In Appendix B, we  discuss some features of multiple Mellin transforms,
detailing the $N=5$ Cayley and $N=6$ Hamilton bases.
In Appendix C, we explicitly perform a quintuple inverse Mellin transform on a generic $5$--point
superstring form factor showing explicitly its result  in terms of delta--functions in Mellin position space.

\section{Unified description of supergravity and superstring amplitudes}

\subsection{Superstring amplitude}
Our discussion builds on the previous studies of the tree-level (disk) superstring amplitudes for the scattering of $N$ gauge bosons, see Refs.
\cite{Stieberger:2006bh,Stieberger:2006te,Stieberger:2007jv,
Stieberger:2007am,Mafra:2011nv,Mafra:2011nw,Stieberger:2012rq}.
In Ref. \cite{Stieberger:2012rq}, specific choices were made for three vertex positions and for the polarization vectors of $N{-}3$ gauge bosons with positive helicities. {}For the partial amplitude associated to the  Tr$(T^{a_1}\cdots T^{a_N})$ Chan-Paton factor, $PSL(2,\IR)$ world-sheet invariance was used to fix $z_1=-\infty, ~z_2=0, ~z_3=1$. Furthermore, the gauge freedom was exercised to select momentum $p_2$ as the reference vector, with the polarization vectors
 \be \varepsilon_k^\mu=\sigma_{a\dot a}^{\mu}\
 {\lambda_2^a\tilde\lambda_k^{\dot a}\over \br 2k\ke}\ ,\quad k\geq 4\ ,
 \ee
 satisfying
 \be
\varepsilon_kp_2=\varepsilon_kp_k=\varepsilon_k\varepsilon_l=0\ .
\ee
The starting point for the MHV superstring formula derived in Ref. \cite{Stieberger:2012rq} is\footnote{Here and later, when discussing MHV graviton amplitudes, we follow the standard practice of omitting the ``supersymmetric'' $\langle 12\rangle^4$ factors associated to two bosons with negative helicities in ``mostly plus'' amplitudes.}
\be A_N^{S}=~\frac{1}{\br 12\ke
\br 23\ke\br 31\ke}\int d\tilde\mu^S_N(z,s)
{\sum_{i\geq 4,~j\geq 3}}^{\!\!\!\prime}~~\prod_{i\neq j}\frac{\varepsilon_{i}p_{j}}{z_{ij}}\ ,\label{samp}\ee
where the integral
\be
\int d\tilde\mu^{S}_N(z,s) := \int_{1}^{\infty}\! dz_4\ldots\int_{z_{N-1}}^{\infty}\!\! dz_N\,
\prod_{2\le k<l\le N}|z_{kl}|^{s_{kl}},\quad s_{kl}=2\alpha'\ p_kp_l\ ,\label{measure}
\ee
and tilde refers to a specific $PSL(2,\IR)$ choice of the three vertex positions, with the  integration domain over the remaining positions correlated with the Chan-Paton factor. Here, as usual, $z_{ij}=z_i-z_j$.
In Eq.(\ref{samp}) the prime over the sum denotes exclusion of any index configuration involving a loop
 $(ij)(jk)\dots (mi)$, thus eliminating all
 closed cycles
$(z_{ij}z_{jk}\cdots z_{mi})^{-1}$ of single poles, in particular the double poles $(z_{ij})^{-2}$ which, upon integration, lead to tachyonic singularities. The remaining $(N-2)^{(N-4)}$ integrals are ``transcendental'' and are characterized by a special form of their small $\alpha'$ (low-energy) expansions.

It is not accidental that, according to Cayley's formula,  $(N-2)^{(N-4)}$ is also the number of tree graphs with $N{-}2$ vertices. In fact, the amplitude (\ref{samp}) can be rewritten as
\be A_N^{S}=~\frac{1}{\br 12\ke\br 23\ke\br 31\ke}\ \bigg(\prod_{k=4}^N\br 2k\ke\bigg)^{-2}\int d\tilde\mu^S_N(z,s)
\sum_{\rm trees}~\prod_{\rm edges}\frac{s_{ij}}{z_{ij}\ z^{\prime}_{ij}}\ ,
\label{samp1}
\ee
where the sum extends over all tree graphs with vertices labelled by $3, 4,\dots,N$. Here, we introduced $z'_{ij}\equiv z'_i-z'_j$ with
\be
z'_i\equiv\frac{\langle ix\rangle}{\langle 2x\rangle\langle 2i\rangle}\ ,
\ee
where $\lambda_x\neq\lambda_2$ is an arbitrary spinor. Indeed, by using Schouten's identity
\be\label{varspinor}
z'_{ij}=z'_i-z'_j=\fc{\langle ij\rangle}{\langle 2i\rangle\langle 2j\rangle}\ ,
\ee
and the $x$--dependence cancels in $z'_{ij}$.
A similar identification between spinor brackets and  free--fermion propagators
on the complex plane has appeared before in \cite{Witten:2003nn}.
Note, that the variables \req{varspinor} satisfy the same partial fraction relations
as the positions variables $z_{ij}$
\be\label{KAPR}
\fc{1}{z_{ij}z_{jk}}+\fc{1}{z_{ik}z_{kj}}+\fc{1}{z_{ij}z_{ki}} ~=~0~=~ \fc{1}{z_{ij}'z_{jk}'}+
\fc{1}{z_{ik}'z_{kj}'}+\fc{1}{z_{ij}'z_{ki}'}\ ,
\ee
reflecting
 \be \fc{\lambda_j}{\vev{ij}\vev{jk}}+\fc{\lambda_k}{\vev{ik}\vev{kj}}+
\fc{\lambda_i}{\vev{ij}\vev{ki}}=0\ ,
\ee
which is the relation underlying Schouten's identity.

In Ref. \cite{Stieberger:2012rq}, we used partial fractioning (\ref{KAPR}) in $z$-variables to rewrite the amplitude (\ref{samp1}) as a sum of chains (Hamiltonian paths) rooted at $i=3$, labeled by $(N-3)!$  permutations $\Pc$ of $4,5, \dots,N$:
\begin{eqnarray}
A_N^S
\comment{&=&
\frac{1}{\br 12\ke
\br 23\ke\br 31\ke}\ \bigg(\prod_{k=4}^N\br 2k\ke\bigg)^{-1}
\int d\tilde\mu^S_N(z,s)
\nonumber \\[2mm] &&\hskip 3ex
\times~\sum_{\Pc}~\frac{1}{z_{34}}\br 2|3|4]~\frac{1}{z_{45}}\br 2|3+4|5]\cdots \frac{1}{z_{(N-1)N}}  \br 2|3+4+\dots+(N-1)|N]\nonumber\\}
&=&\frac{1}{\br 12\ke\br 23\ke\br 31\ke}\
\int d\tilde\mu^S_N(z,s)\ \sum_{\Pc}\ \prod_{k=4}^N\frac{\br 2|3+\ldots+(k-1)|k]}{\br 2k\ke}\ \fc{1}{z_{(k-1)k}}\ .
\label{samp2}
\end{eqnarray}
We refer the reader to Appendix A for a detailed
exposition of Cayley graphs and their reduction to $(N-3)!$ Hamilton graphs subject to partial fraction decomposition on the corresponding rational functions.
 The above  result \req{samp2} can be also expressed as:
\be
A_N^S~=~\frac{1}{\br 12\ke
\br 23\ke\br 31\ke}\ \bigg(\prod_{k=4}^N\br 2k\ke\bigg)^{-2}\
\int d\tilde\mu^S_N(z,s)
\sum_{\Pc}~\prod_{i=4}^{N}\bigg(\frac{1}{z_{(i-1)i}}\sum_{m=3}^{i-1}\frac{s_{mi}}{z'_{mi}}\bigg)\ .
\label{samp3}
\ee
Note that the integrand on the r.h.s. of Eq. (\ref{samp1}) is symmetric under $z\leftrightarrow z'$. Later, we will use partial fractioning in $z'$ instead of $z$, in order to make a direct connection with the general formula \cite{Mafra:2011nv,Mafra:2011nw} for superstring disk amplitudes.
\subsection{Supergravity amplitude}
The tree-level MHV formula for the scattering of $N$ gravitons can be written in many ways \cite{Berends:1988zp,Bern:1998sv,Mason:2008jy,Nguyen:2009jk,Hodges:2012ym,Feng:2012sy}.
In particular, in Ref. \cite{Mason:2008jy}, Mason and Skinner
recast the original formula of Berends, Giele and Kuijf \cite{Berends:1988zp} into the
following form\footnote{More precisely, the formula written below follows from \cite{Mason:2008jy} after a trivial relabelling of graviton indices.}:
\be\label{mason}
A_N^G=\frac{1}{\br 12\ke^2\br 23\ke^2\br 13\ke}\ \sum_{\Pc}\frac{1}{\br 1N\ke}
\prod_{k=4}^N\frac{\br 2|3+\ldots+(k-1)|k]}{\vev{2k}}\ \fc{1}{\vev{(k-1)k}}\ .
\ee
Written in this way, the graviton amplitude bears a striking resemblance to
 the superstring amplitude of Eq. \req{samp2}. Furthermore, we can also make a precise connection between the graphs representing the superstring amplitude  (\ref{samp1}) and the graphs introduced in  Refs. \cite{Nguyen:2009jk,Feng:2012sy} to describe the graviton amplitude. To that end,
it is most convenient to use the Feng--He's version \cite{Feng:2012sy} of Hodges' determinant formula \cite{Hodges:2012ym}.

We begin by redefining
\be
z_i\equiv\frac{\langle iy\rangle}{\langle xy\rangle\langle xi\rangle}~,\qquad
z'_i\equiv\frac{\langle ix\rangle}{\langle yx\rangle\langle yi\rangle}\ ,
\ee
where $\lambda_x$ and $\lambda_y$ are two arbitrary reference spinors.
Furthermore, we have:
\be\label{zij}
z_{ij}=\fc{\langle ij\rangle}{\langle xi\rangle\langle xj\rangle}\ \ \ ,\ \ \
z_{ij}'=\fc{\langle ij\rangle}{\langle yi\rangle\langle yj\rangle}\ ,
\ee
which again obey the partial fraction relations \req{KAPR}. Although $z$ and $z'$ are related by
$z_iz'_i=-\langle xy\rangle^{-2}$, they will be considered as independent variables\footnote{By using the $z$ variables defined  in Eq. \req{zij} with $x=1$, Eq. \req{mason}  may be rewritten as
\be\label{mason1}
A_N^G=\frac{1}{\br 12\ke^2\br 23\ke^2\br 31\ke^2}\ \bigg(\prod_{k=4}^N\br 1k\ke\bigg)^{-2}\ \sum_{\Pc}\
\prod_{k=4}^N\frac{\br 2|3+\ldots+(k-1)|k]}{\vev{2k}}\ \fc{1}{z_{(k-1)k}}\ ,\nonumber
\ee
which assumes the same form as  the superstring amplitude \req{samp2}, where $z$'s denote the vertex positions.}.
Written in terms of these variables, up to an overall sign, the $N$-graviton MHV amplitude \cite{Feng:2012sy} becomes
\be A_N^G= \bigg(\prod_{n=1}^N\br xn\ke\br yn\ke\bigg)^{-2}\frac{1}{z_{ij}z_{jk}z_{ki}}\frac{1}{z'_{rs}z'_{st}z'_{tr}}\left|\Psi\right|^{rst}_{ijk}\ ,
\ee
where $\Psi$ is a $N\times N$ ``weighted Laplacian'' matrix with the elements
\begin{eqnarray} \psi_{ij}=
\left\{  \begin{array}{ll} \displaystyle \frac{s_{ij}}{z_{ij}z'_{ij}} & \textrm{if}~i\neq j\ ,
\\[3ex]\displaystyle  -\sum_{j\neq i}\frac{s_{ij}}{z_{ij}z'_{ij}}  & \textrm{if}~i= j\ ,\end{array}\right.\label{psimatrix}\end{eqnarray}
and $\left|\Psi\right|^{rst}_{ijk}$ denotes the minor determinant obtained after deleting three rows $i,j,k$ and three columns $r,s,t$.

{}For our purposes, it is sufficient to consider the case of $i=r, ~j=s, ~k=t$. Then, according to the matrix--tree theorem, the determinant is given by the sum of all forests consisting of three trees rooted at $i$, $j$ and $k$, with a combined number of $N-3$ edges,  each of them bringing a $\psi_{ij}$ factor. An even simpler expression can be obtained by choosing $x=i,~ y=j$ which sends $z_i\to \infty$ and $z'_j\to \infty$, thus leaving only single trees rooted at $k$. As a result, one obtains all trees with $N-2$ vertices different from $i$ and $j$. {}For example, with the choice $x=i=r=1, ~y=j=s=2, ~k=t=3$, one obtains (\cf \cite{Bern:1998sv,Nguyen:2009jk})
\be A_N^G= \frac{1}{\br 12\ke^2\br 23\ke^2\br 31\ke^2}\
\bigg(\prod_{k=4}^N\br 1k\ke\br 2k\ke\bigg)^{-2}
\sum_{\rm trees}~\prod_{\rm edges}\frac{s_{ij}}{z_{ij}\ z^{\prime}_{ij}}\ ,
\label{gamp1}
\ee
where the sum is over the same trees as in the superstring amplitude (\ref{samp1}).
The edge factors become identical upon reverting to $z$'s defined as vertex positions.
\comment{The gravity amplitude \req{gamp1} has been given in Ref. \cite{Nguyen:2009jk}, see also
\cite{Bern:1998sv}. Similarly, the gravity amplitude
\req{mason} may also be written as
\be\label{mason1}
A_N^G=\frac{1}{\br 12\ke^2\br 23\ke^2\br 31\ke^2}\ \bigg(\prod_{k=4}^N\br 1k\ke\bigg)^{-2}\ \sum_{\Pc}\
\prod_{k=4}^N\frac{\br 2|3+\ldots+(k-1)|k]}{\vev{2k}}\ \fc{1}{z_{(k-1)k}}\ ,
\ee
which assumes the form of the superstring amplitude \req{samp2} upon reverting $z$'s according to \req{zij}, with $x=1$.}
\subsection{Unified description}
On the basis of preceding observations, we can set up a unified description of the amplitudes. First, we define the fundamental function:
\be
{\mathfrak M}_N(z,z',s)^{rst}_{ijk}\ \equiv\
\frac{1}{z_{ij}z_{jk}z_{ki}}\frac{1}{z'_{rs}z'_{st}z'_{tr}}\left|\Psi\right|^{rst}_{ijk}\ .\label{mfunc}
\ee
Then the graviton amplitude is:
\be\label{deltG}
A_N^G\ =\ \int d\mu_N^G(z,\lambda)\int d\mu_N^G(z',\lambda)\
{\mathfrak M}_N(z,z',s)^{rst}_{ijk}\ ,
\ee
with the measures
\begin{eqnarray} \begin{array}{l} \displaystyle
\int d\mu_N^G(z,\lambda)=\int\prod_{i=1}^Ndz_i\ \delta\bigg(\langle xi\rangle^2 z_i-\frac{\langle xi\rangle
\langle yi\rangle}{\langle xy\rangle}\bigg)\ ,
\\[3ex]\displaystyle
\int d\mu_N^G(z',\lambda)=\int\prod_{i=1}^Ndz'_i\ \delta\bigg(\langle yi\rangle^2 z'_i-\frac{\langle xi\rangle
\langle yi\rangle}{\langle yx\rangle}\bigg)\ .
\end{array}\label{measure2}\end{eqnarray}
On the other hand, the superstring amplitude is:
\be
A_N^S\ =\ \int d\mu_N^S(z,s)_{ijk}\int d\mu_N^G(z',\lambda)\ {\mathfrak M}_N(z,z',s)^{rst}_{ijk}\ ,
\label{main}\ee
where:
\be \int d\mu_N^S(z,s)_{ijk}=z_{ij}z_{jk}z_{ki}\int_D\ \bigg(\ {\prod_{l\leq N}}'dz_l\bigg){\prod_{m<n\leq N }\hskip -1.6ex} '\ |z_{mn}|^{s_{mn}}\ .\label{stringm}
\ee
and the primes over products denote exclusion of the indices $i,~j,~k$.
 The domain of integration $D$ along the boundary of the disk is determined by the Chan--Paton factor. The factor $z_{ij}z_{jk}z_{ki}$ in the string measure can be identified as the standard reparametrization ghost correlator.

Finally, we wish to make a comment on the relation between Eq. (\ref{main}) and the general formula \cite{Mafra:2011nv,Mafra:2011nw} for superstring disk amplitudes. Let us start from
Eq. (\ref{main}) with $i=r=N$ and set $z_N\to \infty$, as in Ref. \cite{Mafra:2011nv,Mafra:2011nw}. Next, we choose $y=j=s=N-1$, which sets  $z'_{N{-}1}=\infty$. Finally, we set $k=t=1$. As a result, we obtain all tree graphs with the vertices labeled by $1,2,\dots,N-2$. After partial fractioning in $z'$, in exactly the same way as it was done for $z$ in the derivation of Eq. (\ref{samp3}), we obtain
\be
A_N^S~=~\frac{1}{\br (N{-}1)N\ke
\br N1\ke\br 1(N{-}1)\ke}\ \bigg(\prod_{i=2}^{N-2}\br (N{-}1)i\ke\bigg)^{-2}
\int d\tilde\mu^S_N(z,s)\sum_{\Pc}
\prod_{k=2}^{N- 2}\bigg(\frac{1}{z'_{(k-1)k}}\sum_{m=1}^{k-1}\frac{s_{mk}}{z_{mk}}\bigg)\ ,\label{test}
\ee
where the permutations $\Pc$ are now acting on $2,\dots,N-2$. Here, $z'$ have already been fixed by the $d\mu_N^G(z',\lambda)$ integrations:
\be
z'_i=\frac{\langle iN\rangle}{\langle (N{-}1)N\rangle\langle (N{-}1)i\rangle}\ ,\qquad z'_{ij}=\frac{\langle ij\rangle}{\langle (N{-}1)i\rangle\langle (N{-}1)j\rangle}\ .\ee
After substituting these expressions, Eq. (\ref{test}) becomes
\be
A_N^S~=~
\int d\tilde\mu^S_N(z,s)
\sum_{\Pc}~\bigg(\prod_{k=2}^{N-2}\sum_{m=1}^{k-1}\frac{s_{mk}}{z_{mk}}\bigg)\
A^{Y\!M}_N(1,2,\dots, N)
\ ,\label{test2}
\ee
where $A^{Y\!M}_N$ is the Yang--Mills MHV amplitude \cite{Parke:1986gb}:
\be
A^{Y\!M}_N(1,2,\dots, N)=\frac{1}{\br 12\ke\br 23\ke\cdots\br N1\ke}\ .
\ee
In this way, we obtain the string amplitude in exactly the same form as in Ref.
\cite{Mafra:2011nv}.

%%%%%%%%%%%%%%%%%%%%%%%%%%%%%%%%%%%%%%%%%%%%%%%%%%%%%%%%%%%%%%%%%%%%%%%%%%
\section{String theory  in  Mellin space}
\def\Pc{{\cal P}}
\def\FF#1#2{{_#1F_#2}}

\subsection{Dual resonance models and multi--channel variables}

In dual resonance models\footnote{For a review of dual resonance models,
see Ref. \cite{Frampton}.}, a generic $N$--point scattering process involves multiple resonance exchanges in various channels.
A planar channel  that includes external particles $i,i+1,i+2,\ldots,j-1,j$ is labeled by $(i,j)$.
For a given planar ordering $(1,2,\dots,N)$ of  external particles, there exist $N(N-3)\over 2$
planar channels. The basic property of dual models is that only $N-3$ channels can simultaneously
appear in a given $N$--point (planar) dual diagram.
Actually, not all combinations of channels are allowed. For example,
the adjacent channels like $(i,i+1)$ and $(i+1,i+2)$ cannot appear simultaneously (they are called dual or incompatible channels), while the so--called compatible (non--dual) channels can show up simultaneously\footnote{For precise definitions of compatible and incompatible channels, see
Ref. \cite{DUAL}.}.

An elegant way to account for compatible and incompatible channels
is to
introduce the variables
\be
0\leq u_{i,j}\leq 1\ \ \ \lf\{\begin{array}{ll}
&i=2, ~j=3,\dots, N-1\ ,\\
&i=3,\dots,N-1~<~j=4,\dots,N\ ,
\end{array}\ri.
\label{ijs}
\ee
conjugate to the channels $(i,j)$. Such variables appeared first as the integration variables in the original constructions  of the integral representation of the amplitude generalizing the Veneziano amplitude to an arbitrary number of external particles \cite{DUAL}.
In the following, the above set \req{ijs} of indices $(i,j)$  is denoted by $P$.
There are $N(N-3)\over 2$ such coordinates $u_{i,j}$, in one--to--one correspondence with the independent\footnote{Here, we ignore constraints related to finite dimensionality of space--time, see \cite{Stieberger:2007jv}.} kinematic invariants
\be\label{mandelstam}
s_{i,j}=\ap (k_i+k_{i+1}+\ldots+k_j)^2
\ee
associated to the allowed planar channels $(i,j)$ of the $N$--point scattering amplitude.
For a given channel $(i,j)$ with $u_{i,j}=0$, all incompatible channels $(k,l)$ are required to have $u_{k,l}=1$.
These conditions  can be summarized by the following $N(N-3)\over 2$ (nonlinear) constraints
\be\label{constraints}
u_P=1-\prod_{\widetilde P}u_{\widetilde P}\ ,
\ee
with $\widetilde P$ the set of all channels incompatible to $P$. This set of equations is sufficient
for excluding simultaneous poles in incompatible channels.
Only $(N-2)(N-3)\over 2$ of the above constraints \req{constraints}
are independent, thus leaving $N-3$ free variables which can be chosen as
$u_{2,j},\ j=3,\ldots,N-1$. In this way, a generic $N$-particle  dual amplitude can be written as
\be\label{FORM}
B_N(\{s_{k,l}\},\{n_{k,l}\}):=\lf(\prod_{i,j\in P}\int_0^\infty  du_{i,j}\ u_{i,j}^{s_{i,j}-1+n_{i,j}}\ \theta(1-u_{i,j})\ri)\ \delta(\{u_{p,q}\})\ ,
\ee
with a set $n_{i,j}$ of integers and the product of  $(N-2)(N-3)\over 2$  delta functions
\be\label{olddelta}
\delta(\{u_{p,q}\})={\prod_{P}}^{\,\prime}
\delta\lf(u_{P}-1+\prod_{\widetilde P}u_{\widetilde P}\ri)\ ,
\ee
where the prime over product indicates exclusion of the channels $(2,j)$.

While Quantum Chromodynamics superseded the dual resonance model as the theory of strong interactions,
all assumptions/axioms of duality have been later implemented in disk amplitudes of open superstring theory. Actually, the change of integration variables from the positions of vertex operators at the disk boundary to the multi--channel variables (\ref{ijs}) is very useful for studying the singularity structure of the amplitudes \cite{Mafra:2011nw} and allows rewriting a generic string ``formfactor" in exactly the same form as the dual amplitude (\ref{FORM}). This change of integration variables is described below.

\subsection{Pascal's triangle of constraints}

Here, we give another representation of the constraints \req{constraints} which is more natural for the computations of open string disk amplitudes. Now  the $N(N-3)\over 2$ coordinates (\ref{ijs}) are written as the M\"obius--invariant cross--ratios of vertex positions at the boundary
\be
u_{i,j}={(z_i-z_j)\ (z_{i-1}-z_{j+1})\over(z_i-z_{j+1})\ (z_{i-1}-z_j)}\ ,
\label{crossr}
\ee
with the indices $i,j\in P$ specified in (\ref{ijs}) and  the cyclic identification $k+N\equiv k$ (\eg\ $z_0=z_N,~z_{N+1}=z_1$ etc.).
By using elementary algebraic manipulations, it is easy to show that these coordinates do indeed satisfy Eq. (\ref{constraints}).

The new set of (equivalent) constraints can be succinctly summarized by drawing a ``Pas{\nolinebreak}cal's triangle'' of $(N-2)(N-3)\over 2$ cells labeled by $(k,l)$, $k=3,\dots,N-1$ $<$ $l=4,\dots,N$ or equivalently, by the sequences $(k,k+1,\dots, l)$, see Figure~1. \vskip 4ex
\begin{center}
\fcolorbox{white}{white}{
  \begin{picture}(364,340) (147,-75)
    \SetWidth{1.0}
    \SetColor{Black}
    \Line(336,244)(354,214)
    \Line(336,244)(318,214)
    \Text(258,172)[lb]{\LARGE{\Black{${\displaystyle(3,N{-}1)\atop
    \scriptscriptstyle (3,4,\dots,N{-}1)}$}}}
    \Text(348,172)[lb]{\LARGE{\Black{${\displaystyle(4,N)\atop\scriptscriptstyle (4,5\dots,N)}$}}}
    \Text(312,244)[lb]{\LARGE{\Black{${\displaystyle(3,N)\atop\scriptscriptstyle (3,4,\dots,N)}$}}}
    \Line(294,172)(276,142)
    \Line(378,172)(396,142)
    \Line(354,172)(336,142)
    \Line(318,172)(336,142)
    \Text(385,106)[lb]{\LARGE{\Black{${\displaystyle(5,N)\atop\scriptscriptstyle (5,6,\dots,N)}$}}}
    \Text(303,106)[lb]{\LARGE{\Black{${\displaystyle(4,N{-}1)\atop\scriptscriptstyle (4,5\dots,N{-1})}$}}}
    \Text(222,106)[lb]{\LARGE{\Black{${\displaystyle(3,N{-}2)\atop\scriptscriptstyle (3,4,\dots,N{-}2)}$}}}
    \Line(186,-14)(168,-44)
    \Line(258,106)(240,76)
    \Line(492,-14)(510,-44)
    \Line(420,106)(438,76)
    \Line[dash,dashsize=2.6](264,94)(414,94)
    \Text(442,-8)[lb]{\LARGE{\Black{${\displaystyle(N{-}2,N)\atop\scriptscriptstyle (N{-}2,N{-}1,N)}$}}}
    \Text(180,-8)[lb]{\LARGE{\Black{${\displaystyle(3,5)\atop\scriptscriptstyle (3,4,5)}$}}}
    \Line[dash,dashsize=2.6](438,76)(468,28)
    \Line[dash,dashsize=2.6](240,76)(210,28)
    \Line(210,-14)(228,-44)
    \Line(306,-14)(288,-44)
    \Line(270,-14)(288,-44)
    \Line(246,-14)(228,-44)
    \Line[dash,dashsize=2.6](320,-24)(480,-24)
    \Text(240,-8)[lb]{\LARGE{\Black{${\displaystyle(4,6)\atop\scriptscriptstyle (4,5,6)}$}}}
    \Text(300,-8)[lb]{\LARGE{\Black{${\displaystyle(5,7)\atop\scriptscriptstyle (5,6,7)}$}}}
    \Text(270,-70)[lb]{\LARGE{\Black{${\displaystyle(5,6)}$}}}
    \Text(210,-70)[lb]{\LARGE{\Black{${\displaystyle(4,5)}$}}}
    \Text(144,-70)[lb]{\LARGE{\Black{${\displaystyle(3,4)}$}}}
    \Text(475,-70)[lb]{\LARGE{\Black{${\displaystyle(N{-}1,N)}$}}}
    \end{picture}}
\end{center}
\begin{center}FIG.1. Pascal's triangle of constraints.
\end{center}
\addtocounter{figure}{1}
Each cell has its ancestors at higher levels and descendants at lower levels, sharing a sequence of at least two indices. For example, in Figure 1, the cell $(4,N{-}1)$ has 3 ancestors:
$(3,N{-}1), (4,N), (3,N)$, while the ancestors of $(3,4)$ are: $(3,5),(3,6),\dots,(3,N)$. Note that $(3,N)$, at the top of the triangle, is the primary ancestor to all cells.
{}For each cell $(k,l)$, we define the homogenous functions
\be \rho_{kl}(u)=u_{k,\, l}\prod_{a}a(u_{k,\, l})\qquad,\qquad \sigma_{kl}(u)=\prod_{n=k}^{l-1}u_{2,n}\ ,
\ee
where $a(u_{k,\,l})$ are the ancestors of $u_{k,\,l}$.
We also introduce the polynomials:
\be
\alpha_{kl}(u) =\rho_{kl}(u)+ \sigma_{kl}(u)-1 \ .
\label{alo}
\ee
To each cell of the triangle, we associate the constraint:
\be
\alpha_{kl}(u)=0\ . \label{uconst}
\ee
Here again, it is a matter of simple algebra to verify these constraints and to show that that they are equivalent to Eq. (\ref{constraints}), thus describing the same embedding of disk boundary in $N(N-3)\over 2$--dimensional projective space.

\subsection{From string world--sheet to Mellin space}

We wish to uplift the integrals (\ref{measure}) over string vertex positions $z_1,z_2,\dots,z_N$ from disk boundary to the projective space of conformal coordinates (\ref{crossr}). The constraints (\ref{uconst}) will be implemented by inserting the following product of delta functions
\be\label{deltaM}
\delta(\{u_{p,q}\})
=\prod_{l=4}^{N}\delta(\alpha_{l-1,l})\ \prod_{k=3}^{l-2}\rho_{kl}\ \delta(\alpha_{kl})\ ,
\ee
which is equivalent to \req{olddelta}. In this way, the string integral measure  (\ref{measure}) is replaced by the following integral:
\bea\label{zmel}
\int d\tilde\mu^{S}_N(z,s)&=&\int d M_N(u,s)\\
&:=&\lf(\prod_{i,j\in P}\int_0^\infty  du_{i,j}\ u_{i,j}^{s_{i,j}-1}\ \theta(1-u_{i,j})\ri) \
\lf(\prod_{l=4}^{N}\sigma_{3l}^{-1}\ \rho_{l-1,l}\ri)\ \delta(\{u_{p,q}\})\ .\nonumber\eea
Here, the  delta functions enforce the constraints (\ref{uconst}), while  the adjacent bracket  comprises
a Jacobian determinant, which follows from the differential\footnote{Note the identity:
$\fc{z_{12}z_{23}z_{31}}{\prod\limits_{l=1}^N|z_l-z_{l+2}|}=\det\lf(\fc{\p u_{p,q}}{\p z_{mn}}\ri)\
\prod\limits_{2\leq i<j\leq N-1} u_{N+2-j,N+2-i}^{1-j+i}=
\lf(\prod\limits_{l=4}^{N}\sigma_{3l}\ \rho_{l-1,l}^{-1}\ri)\ \lf(\prod\limits_{i,j\in P}u_{i,j}\ri)\ .$}:
\be\label{DIFFERENTIAL}
\lf(\prod_{k=4}^N dz_k\ri)\ \lf(\prod_{i<j}^N |z_{ij}|^{s_{ij}}\ri)\
\fc{z_{12}z_{23}z_{31}}{\prod\limits_{l=1}^N|z_l-z_{l+2}|}
=\lf(\prod_{i,j\in P} du_{i,j}\ u_{i,j}^{s_{i,j}}\ri)\ \delta(\{u_{p,q}\})\ .
\ee
Written explicitly in terms of the delta functions associated to the cells of Pascal's triangle,
\be
\int d M_N(u,s)=\lf(\prod_{i,j\in P}\int_0^\infty  du_{i,j}\ u_{i,j}^{s_{i,j}-1}\ \theta(1-u_{i,j})\ri) \
\lf(\prod_{l=4}^{N}\sigma_{3l}^{-1}\prod_{k=3}^{l-1}\ \rho_{kl}\,\delta(\alpha_{kl})\ri)\ .\label{zmmeli}
\ee
The above integral represents a multi--dimensional Mellin transform\footnote{The (single variable) Mellin transformation is an operation $M$, which assigns a function $M_f(s):=\int\limits_0^\infty dx\ x^{s-1}\ f(x)$ of the complex variable $s$ to each locally
summable function $f(x)$, which satisfies the following two conditions:
$(i)$ $f(x)$ is defined for $x>0$ and $(ii)$ there exists a strip $s_1<\Re(s)<s_2$ in the complex $s$--plane such that $x^{s-1} f(x)$ is absolutely integrable w.r.t. $x\in (0,\infty)$.}, directly from the string world--sheet boundary to the dual space of kinematic invariants $s_{i,j}$, called Mellin space, thus side--stepping space--time.

In the previous Section we concluded that both the (MHV) scattering amplitudes $A_N^S$ of gauge bosons in superstring theory and the graviton amplitudes $A_N^G$ of supergravity can be obtained from a single function $\mathfrak{M}_N$, \cf Eq. (\ref{mfunc}), by appropriate integrations as \req{deltG} and \req{main}, respectively.
Thus we can start from the supergravity amplitude \req{deltG} and replace the integral measure
$\int d\mu^{G}_N(z,\lambda)$  by the Mellin transform (\ref{zmel})
\be\label{duality}
\int d\mu^{G}_N(z,\lambda)\to \int dM_N(u,s)\ ,
\ee
which, in the specific gauge, takes us back to the tree--graph formula of
Eq. (\ref{samp1}) with
\be\label{duality1}
\int d\tilde\mu^{S}_N(z,s)\to\int dM_N(u,s)
\ee
and the edge factors expressed in terms of the  position variables $u_{i,j}$ described in the following.

As written in Eq. (\ref{samp1}), in addition to the position variables, the edge factors
$\fc{s_{ij}}{z_{ij}}$ depend on  the kinematic invariants $s_{i,j}$ belonging to the dual Mellin space. Through $z_{ij}^{s_{ij}}\fc{s_{ij}}{z_{ij}}=\p_{z_{ij}}z_{ij}^{s_{ij}}$, these can be represented by
insertions of the respective differential operators acting in
position space $\p_{u_{i,j}} u_{i,j}^{s_{i,j}}$.
After integrating by parts, their action can be redirected on the delta function constraint
\req{deltaM}.
One finds that each edge gives rise to a single derivative of the respective delta function
\be\label{ruledelta}
\fc{s_{ij}}{z_{ij}}\to-\sigma_{3j}\ \fc{\delta'(\alpha_{ij})}{\delta(\alpha_{ij})}\ ,
\ee
supplemented by the factor $\sigma_{3j}$.
Eventually\footnote{We refer the reader to Appendix B for further technical details and examples.}, by using the relation $x\;\delta'(x)=-\delta(x)$ \cite{Vladimirov}, the replacements
\req{duality1} and \req{ruledelta} provide the final form of the amplitude (\ref{samp1}):
\be A_N^{S}=~\frac{1}{\br 12\ke\br 23\ke\br 31\ke}\ \bigg(\prod_{k=4}^N\br 2k\ke\bigg)^{-2}
\ \int dM_N(u,s)\ \sum_{\rm trees}~\prod_{\rm edges}\
\frac{\sigma_{3j}}{\alpha_{ij}\, z^{\prime}_{ij}}\ .
\label{smel1}
\ee
Here, the sum extends over all tree graphs with vertices labeled by $i,j=3, 4,\dots,N$, as in
Eq.~(\ref{samp1}), and we used the fact that graphs are unoriented to label edges by ordered pairs $i<j$.
The above formula yields the superstring amplitude as a Mellin transform of the graviton supergravity amplitude \req{gamp1}.

\comment{\be\label{FINALM}
A_N^S=\vev{12}\vev{23}\vev{31}\ \lf(\prod_{k=4}^N\vev{1k}\ri)^2\
\int dM_N(u,s)\  \lf.A_N^G\ \ri|_{\fc{s_{ij}}{z_{ij}}\ra\sigma_{3j}\
\p\ln\delta(\alpha_{ij})}\ .
\ee}

\section{Superstring amplitudes as Mellin amplitudes}

In quantum field theory, generic correlation functions
do not assume a simple form in position space, therefore one performs a Fourier transform to momentum space, where the analytic properties of the amplitude like its pole structure become simple. For CFTs, the
Symanzik's star operator
converts  position space integrals into  (inverse) Mellin transforms of the so--called
Mellin amplitudes \cite{Symanzik:1972wj}. The latter depend on complex variables
$s_{i,j}$ substituting for the kinematic invariants of the scattering amplitude. Hence for CFTs, a Mellin transform
is more appropriate and Mellin space serves as a natural momentum space.
In this space, the CFT amplitudes exhibit a universal behaviour.
Thus Mellin transforms, although less familiar to  particle theorists than Fourier transforms\footnote{In fact, the theory of Mellin transform is equivalent to that of Fourier transforms in the complex plane \cite{Tit}: For the Mellin transform $M_f(s)$ of a function $f$ with
$s\in \IC$ we have
$(2\pi)^{-1/2}\ M_f(is)=(2\pi)^{-1/2}\int\limits_{-\infty}^\infty\ dy\ e^{-isy}\ g(y)$, with the latter being the Fourier transform of $g(y):=f(e^y)$.}, are useful for representing correlation functions in CFTs, so it is not too surprising that they appear in the context of the two--dimensional string world--sheet.
In this Section we want to investigate the connection between generic dual (superstring) amplitudes
\req{FORM} and Mellin amplitudes in more detail.

\subsection{Mellin amplitudes}

In Refs. \cite{Mack,Mack1} an exact correspondence between conformal field theories in $D$ dimensions and
dual resonance models in $D'$ dimensions has been established. Correlation functions
in the $D$--dimensional CFT are related to Mellin amplitudes of  the dual resonance model \cite{Veneziano:1968yb} through an inverse Mellin transformation\footnote{The (single variable) inverse Mellin transform $f$ of a function $M_f(s)$ is given by the complex integral
$f(x)=\fc{1}{2\pi i}\int\limits_{-i\infty+c}^{+i\infty+c} ds\ x^{-s}\ M_f(s)$,
provided that  the latter converges absolutely along the line $s=c$ for any real value $c$ with $s_1<c<s_2$, the function
$M_f(s)$ is analytic in the strip $s_1<\Re(s)<s_2$ and  goes to zero uniformly for increasing $\Im(s)$.}.
The latter relates a  conformal $N$--point  function
$\Gc_N(\{x_r\}):=\Gc(x_1,\ldots,x_N)$ with positions $x_r$ in $D$ dimensions
and  depending on $m=\h N(N-3)$  anharmonic cross--ratios
$\omega_{i,j}$  to a scattering amplitude $\Mc(\{s_{k,l}\})$ in the
dual resonance model in $D'$ dimensions as
\be\label{Mellin}
\Gc_N(\{x_r\})=(2\pi i)^{-m}\ \lf(\prod_{i,j\in P}\int_{-i\infty+c}^{+i\infty+c} ds_{i,j}\ \omega_{i,j}^{-s_{i,j}}\ri)\ \lf(\prod_{i<j}^N\Gamma(s_{ij})\ri)\ \Mc(\{s_{k,l}\})\ ,
\ee
with $m$ complex variables $s_{i,j}$ to be related to the $m$ kinematic invariants  \req{mandelstam} and the cross--ratios $\omega_{i,j}={(x_i-x_j)(x_{i-1}-x_{j+1})\over(x_i-x_{j+1})(x_{i-1}-x_j)}$ for $D=2$ and $\omega_{i,j}={|x_i-x_j|^2|x_{i-1}-x_{j+1}|^2\over|x_i-x_{j+1}|^2|x_{i-1}-x_j|^2}$ for $D>2$.
In \req{Mellin} the integration is over a suitable choice of contour in the complex variables
$s_{i,j}$ and the set of indices $i,j$ is defined in \req{ijs}.
The relation \req{Mellin} has recently been applied to conformal theories in $D=4$ dimensions to derive conformal
correlators in AdS/CFT backgrounds from Mellin space \cite{Fitzpatrick:2011ia}, to find Feynman  rules for  Mellin amplitudes  \cite{Paulos:2011ie}, and to rewrite  dual conformal integrals of perturbative scattering amplitudes in ${\cal N}=4$ SYM~\cite{Paulos}.

Furthermore, it has been speculated in \cite{Mack}, that the  Mellin amplitude
$\Mc(\{s_{k,l}\})$, which shares exact duality, \ie meromorphy in $s_{i,j}$ with simple poles in single variables, crossing symmetry and factorization, may actually be derived from correlators in string theory.
In fact,  in the following\footnote{In the sequel  we shall work with the reduced Mellin amplitude
$\hat{\Mc}(\{s_{k,l}\})=\Mc(\{s_{k,l}\})\ (\prod\limits_{i<j}^N\Gamma(s_{ij}))$ \cite{Mack1}.} we shall start from the specific $N$--point Mellin amplitudes
\be\label{idneti}
\widehat \Mc(\{s_{k,l}\})=B_N(\{s_{k,l}\},\{n_{k,l}\})
\ee
describing dual (superstring) $N$--point amplitudes \req{FORM}
and consider their inverse Mellin transforms \req{Mellin}. At a practical level
in this case the objects $\widehat\Mc(\{s_{k,l}\})$ represent complicated multiple Gaussian hypergeometric functions encoding the infinite heavy string states and their inverse
Mellin transforms should give some simple function $\Gc(\{x_r\})$ describing
correlators of a conformal field theory.

Finally, integral transformations
% such as Sommerfeld--Watson transformations or
on string form factors \req{FORM}
have already been considered in the past for computing high--energy
limits,  dispersion relations and discontinuities of dual amplitudes. {\it E.g.}
in Ref.~\cite{Veneziano:1974dr}
 a dual amplitude $B_N$, as a function of its own set of (planar) kinematic invariants
$s_{i,j}$, is written
as a multiple beta--transform on some conjugate set of the variables $s_{k,l}$.
For $N=4$ this transformation gives rise to \cite{Coral}
$$B(s,u)=
\int\limits_{-i\infty+c}^{+i\infty+c}d\sigma\int\limits_{-i\infty+c}^{+i\infty+c}d\tau\
B(s,\sigma)\ B(u,\tau)\ B(1-\sigma,1-\tau)\ ,$$
with $s=s_{1,2},\ u=s_{2,3}$ and the Euler beta function
$B(s,u)=\fc{\Gamma(s)\Gamma(u)}{\Gamma(s+u)}$.
However, as we will see in the next paragraph multiple Mellin transforms and its application to
distributional delta--functions provide a novel direction.

\subsection{String form factors, inverse Mellin transforms and space--time correlators}

According to Section 3 for a set of integers $n_{i,j}$ the form factors \req{FORM}
of the $N$--point string amplitude can be written in terms of integrals over the $m:=\h N(N-3)$ coordinates \req{ijs}
with the product of delta functions $\delta(\{u_{k,l}\})$
given in \req{olddelta} or \req{deltaM}.
We refer the reader to Ref. \cite{Mafra:2011nw} for a detailed exposition\footnote{Note, that the coordinates \req{ijs} are subject to the identifications
$u_{i,j}=u_{j+1,i-1}$ and $u_{k,N}=u_{1,k-1},\ k\geq 3$.}
and application of \req{FORM} in view of the hypergeometric function
structure of superstring amplitudes. {\it E.g.} for $N=4$ the expression \req{FORM} gives
\bea
B_4(\{s_{k,l}\},\{n_{k,l}\})&=&\int_0^1 du_{1,2}\ \int_0^1 du_{2,3}\ u_{1,2}^{s_{12}-1+n_{1,2}}\ u_{2,3}^{s_{23}-1+n_{2,3}}\
\delta(u_{1,2}+u_{2,3}-1)\nnn
&=&\fc{\Gamma(s+n_{1,2})\ \Gamma(u+n_{2,3})}{\Gamma(s+u+n_{1,2}+n_{2,3})}\ ,\label{N4}
\eea
while for $N=5$ we have
\bea
B_5(\{s_{k,l}\},\{n_{k,l}\})
&=&\int_0^1 du_{1,2}\int_0^1 du_{2,3}\int_0^1 du_{3,4}\int_0^1 du_{4,5}
\int_0^1 du_{1,5}\ u_{1,2}^{s_1-1+n_{1,2}}\ u_{2,3}^{s_2-1+n_{2,3}}\ u_{3,4}^{s_3-1+n_{3,4}}\nnn
&\times& u_{4,5}^{s_4-1+n_{1,3}}\ u_{1,5}^{s_5-1+n_{2,4}}\ \
\delta(u_{1,2}+u_{2,3}u_{1,5}-1)\ \delta(u_{3,4}+u_{2,3}u_{4,5}-1)\nnn
&\times&\ds{\delta(u_{4,5}+u_{3,4}u_{1,5}-1)=
\fc{\Gamma(s_2+n_{2,3})\ \Gamma(s_3+n_{3,4})}{\Gamma(s_2+s_3+n_{2,3}+n_{3,4})}
\fc{\Gamma(s_4+n_{4,5})\ \Gamma(s_5+n_{1,5})}{\Gamma(s_4+s_5+n_{4,5}+n_{1,5})}}\nnn
&\times&\ds{\FF{3}{2}\lf[{s_2+n_{2,3},\ s_5+n_{1,5},\ s_3+s_4-s_1+n_{3,4}+n_{4,5}-n_{1,2}\atop
s_2+s_3+n_{2,3}+n_{3,4},\ s_4+s_5+n_{4,5}+n_{1,5}};1\ri]}\ ,\label{N5}
\eea
with\footnote{Note, that the integers $n_{i,j}$ are subject to the identifications
$n_{i,j}=n_{j+1,i-1}$ and $n_{k,N}=n_{1,k-1},\ k\geq 3$, \ie $n_{1,3}=n_{4,5},\ n_{1,5}=n_{2,4}$ and
$n_{3,5}=n_{1,2}$.}  $s_1=s_{1,2},\ s_2=s_{2,3},\ s_3=s_{3,4},\ s_4=s_{4,5}$ and $s_5=s_{5,1}$.

Let us now compute the integral \req{Mellin} (with the reduced Mellin amplitude \req{idneti})
for the two amplitudes \req{N4} and \req{N5},  respectively.
For $N=4$ we perform the  double inverse Mellin transformation on \req{N4}:
\bea
&&\fc{1}{(2\pi i)^2}\
\int\limits_{-i\infty+c}^{+i\infty+c}ds\int\limits_{-i\infty+c}^{+i\infty+c}du\ u_{1,2}^{-s}\
u_{2,3}^{-u}\ \fc{\Gamma(s+n_{1,2})\ \Gamma(u+n_{2,3})}{\Gamma(s+u+n_{1,2}+n_{2,3})} \nnn
&&\hskip1cm=u_{1,2}^{n_{1,2}}\ u_{2,3}^{n_{2,3}}\ \ \delta(1-u_{1,2}-u_{2,3})
\ \theta(1-u_{1,2})\ \theta(1-u_{2,3})\ .\label{double}
\eea
To show \req{double} we have used \cite{Kaminski}
\be\label{KAMI}
\fc{1}{2\pi i}\ \int_{-i\infty+c}^{+i\infty+c}\ ds\ x^{-s}\ \fc{\Gamma(s)}{\Gamma(s+a+1)}
=\begin{cases}
\fc{\ (1-x)^a}{\Gamma(a+1)}&,\ 0<x\leq 1\ ,\\
\ 0&,\ x>1\ ,
\end{cases}
\ee
for $\Re(a)>-1$ and the inverse Mellin transformation
\be\label{KAMII}
\delta(x-y)=\delta(y-x)=\fc{1}{2\pi i}\int_{-i\infty+c}^{+i\infty+c}\ ds\ x^{-s}\ y^{s-1}\ \ \ ,\ \ \ x,y>0
\ee
following from the Mellin transformation of the $\delta$--function $y^{s-1}\ \theta(y)=\int_0^\infty\ dx\ x^{s-1}\ \delta(x-y)$ for $y>0$ \cite{Kang}.
The extension of the Mellin transformation to  a larger framework, in which Dirac delta and other singular functions can be treated, has mainly been established by Kang \cite{Kang}.
For $N=5$ we must consider the quintuple inverse Mellin transformation on \req{N5}:
\bea
&&\fc{1}{(2\pi i)^5}\
\int\limits_{-i\infty+c}^{+i\infty+c}ds_1\int\limits_{-i\infty+c}^{+i\infty+c}ds_2
\int\limits_{-i\infty+c}^{+i\infty+c}ds_3\int\limits_{-i\infty+c}^{+i\infty+c}
ds_4\int\limits_{-i\infty+c}^{+i\infty+c}ds_5\
u_{1,2}^{-s_1}\ u_{2,3}^{-s_2}\ u_{3,4}^{-s_3}\ u_{4,5}^{-s_4}\ u_{1,5}^{-s_5}\nnn
&&\hskip1cm\ds{\times\fc{\Gamma(s_2+n_{2,3})\
\Gamma(s_3+n_{3,4})}{\Gamma(s_2+s_3+n_{2,3}+n_{3,4})}\
\fc{\Gamma(s_4+n_{4,5})\ \Gamma(s_5+n_{1,5})}{\Gamma(s_4+s_5+n_{4,5}+n_{1,5})}}\nnn
&&\hskip1cm\ds{\times\FF{3}{2}\lf[{s_2+n_{2,3},\ s_5+n_{1,5},\
s_3+s_4-s_1+n_{3,4}+n_{4,5}-n_{3,5}\atop
s_2+s_3+n_{2,3}+n_{3,4},\ s_4+s_5+n_{4,5}+n_{1,5}};1\ri]}\label{QUINT}\\[5mm]
&&\hskip1cm=u_{1,2}^{n_{1,2}}\ u_{2,3}^{n_{2,3}}\ u_{3,4}^{n_{3,4}}\ u_{4,5}^{n_{1,3}}
\ u_{1,5}^{n_{2,4}}\ \ \delta(u_{4,5}+u_{3,4}u_{1,5}-1)\ \delta(u_{1,2}+u_{2,3}u_{1,5}-1)\nnn
&&\hskip1cm\times \delta(u_{3,4}+u_{2,3}u_{4,5}-1)\ \theta(1-u_{1,2})\ \theta(1-u_{2,3})\ \ \theta(1-u_{3,4})\ \theta(1-u_{4,5})\ \theta(1-u_{1,5})\ .\nonumber
\eea
The details of these integrations are displayed in Appendix C.

The two examples $N=4$ and $N=5$ demonstrate, that the inverse Mellin transformation
\req{Mellin} of the string form factors \req{FORM} essentially picks up the delta--functions describing the duality constraint equations \req{uconst}.
Hence, for any $N$ the (multiple) inverse Mellin transform of \req{FORM} yields:
\bea\label{FORMB}
G_N(\{u_{i,j}\})&:=&(2\pi i)^{-m}\ \lf(\prod_{(i,j)\in P}\  \int_{-i\infty+c}^{+i\infty+c} ds_{i,j} \ u_{i,j}^{-s_{i,j}}\ri)\ B_N(\{s_{k,l}\},\{n_{k,l}\})\nonumber\\
&=&\lf(\prod_{(i,j)\in P}\ u_{i,j}^{n_{i,j}}\  \theta(1-u_{i,j})\ri)\
\delta(\{u_{k,l}\})\ .
\eea

It is interesting to look at the analogue of the $\ap$--expansion in  Mellin position space.
{\it E.g.} for $N=4$ Eq. \req{samp1} becomes:
\bea
A_4^S&=&
\fc{1}{
\vev{12}\vev{31}\vev{34}\vev{24}}\ \int d\tilde\mu^S_4(z,s)\ \frac{s_{34}}{z_{34}}
\nonumber\\[3mm]
&=&\fc{1}{
\vev{12}\vev{31}\vev{34}\vev{24}}\ s\ \fc{\Gamma(s)\ \Gamma(u)}{\Gamma(s+u)}\ .
\eea
A double inverse Mellin transform on
\be
s\ \fc{\Gamma(s)\ \Gamma(u)}{\Gamma(s+u)}=1+\fc{s}{u}-\zeta(2)\ (s^2+su)+\Oc(\ap^3)
\ee
yields the corresponding relation in Mellin position space:
\bea
&&u_{1,2}\ \delta'(1-u_{1,2}-u_{2,3})\
\theta(1-u_{1,2})\ \theta(1-u_{2,3})\\[3mm]
&&\hskip0.5cm=\theta(1-u_{1,2})\ \theta(1-u_{2,3})\ \big\{\ \delta(1-u_{1,2})\ \delta(1-u_{2,3})+u_{1,2}\ \delta'(1-u_{1,2})
\nonumber\\[3mm]
&&\hskip0.5cm-\zeta(2)\ [\ u_{1,2}^2\ \delta''(1-u_{1,2})\ \delta(1-u_{2,3})
+u_{1,2}\ \delta'(1-u_{1,2})\ \delta'(1-u_{2,3})\ ]\  \big\}+\ldots\ .
\ \nonumber
\eea
It is straightforward to derive similar expansions for $N\geq 5$. It would be interesting
to relate the  amplitudes \req{FORMB} in position space with the results in \cite{KLT},
\ie relating aspects of motivic multiple zeta values to the space of Dirac delta and other singular functions.

\subsection{Mellin position space and conformal cross--ratios}

According to \req{crossr} the $m$ channel variables $u_{i,j}$ can be identified with the  anharmonic ratios, \ie $\omega_{i,j}=u_{i,j}={(z_i-z_j)\ (z_{i-1}-z_{j+1})\over(z_i-z_{j+1})\ (z_{i-1}-z_j)}$.
With this choice  all duality constraint equations \req{uconst} are satisfied reducing
the $m$ anharmonic ratios to $N-3$ independent fundamental cross--ratios. Rewriting \req{FORM}
in terms of \req{crossr} reveals the invariance group $PSL(2,\IR)$.
{\it E.g.} for $N=4$ the two variables $u_{1,2}$ and $u_{2,3}$ are identified as
\be
u_{1,2}=\fc{z_{12}z_{34}}{z_{13}z_{24}}\ \ \ ,\ \ \ u_{2,3}=\fc{z_{23}z_{14}}{z_{13}z_{24}}\ ,
\ee
with:
\be
u_{2,3}=1-u_{1,2}\ .
\ee
Furthermore, for $N=5$ we have the following relations
\ba\label{u5}
u_{1,2}&=&\fc{z_{12}z_{35}}{z_{13}z_{25}}\ \ \ ,\ \ \
u_{2,3}=\fc{z_{23}z_{14}}{z_{13}z_{24}}\ \ \ ,\ \ \
u_{3,4}=\fc{z_{34}z_{25}}{z_{24}z_{35}}\nnn
u_{4,5}&=&\fc{z_{13}z_{45}}{z_{35}z_{14}}\ \ \ ,\ \ \
u_{5,1}=\fc{z_{24}z_{15}}{z_{14}z_{25}}\ ,
\ea
with:
\be
u_{1,2}=1-u_{2,3}u_{1,5}\ \ \ ,\ \ \ u_{3,4}=\fc{1-u_{2,3}}{1-u_{2,3}u_{1,5}}\ \ \ ,\ \ \
u_{4,5}=\fc{1-u_{1,5}}{1-u_{2,3}u_{1,5}}\ .
\ee

The $N$ real variables $z_i$ are associated to each external leg $i$.
It has been known since the early days of superstring theory that the integrals over vertex positions on a disk boundary can be replaced by a different set of variables. From the set $z_1,\dots z_N$ one picks three positions, say  $z_a,z_b,z_c$ and and employs $PSL(2,\IR)$ invariance setting them to specific values. For instance, we can choose
$z_1=-\infty,~z_2=0,~z_3=1$ as in  Section 2.
With \req{crossr} and \req{DIFFERENTIAL} the amplitude \req{FORM} takes the form \cite{KOBA}
\be\label{koba}
B_N(\{s_{k,l}\},\{n_{k,l}\})=\int_{-\infty}^{\infty}\fc{\prod\limits_{i=1}^N\theta(z_i-z_{i+1})\ dz_i}{dz_adz_bdz_c}\ \fc{z_{ab}z_{bc}z_{ca}}{\prod\limits_{i=1}^N |z_i-z_{i+2}|}\
\lf(\prod_{i<j}^N |z_{ij}|^{s_{ij}}\ z_{ij}^{\widetilde n_{ij}}\ri)\ ,
\ee
with:
\be
\widetilde n_{ij}=
n_{i,j}+n_{i+1,j-1}-n_{i+1,j}-n_{i,j-1}\ .
\ee

\section{Concluding remarks}

In this work, we argued that in the case of MHV helicity configurations, $N$--gluon superstring amplitudes are given by Mellin transforms of $N$--graviton supergravity amplitudes, as written in
Eq.\req{smel1}. The most pressing question is whether this result can be extended to other helicity configurations, at least at the semi--classical level, that is promoted to a general relation between all string disk amplitudes and tree--level graviton amplitudes of supergravity.
In principle, all necessary ingredients are available for answering this question: superstring amplitudes are written in
Refs. \cite{Stieberger:2007am} and \cite{Mafra:2011nv,Mafra:2011nw} while (very plausible conjectures for) supergravity amplitudes can be found in Refs. \cite{Cachazo:2012da,Cachazo:2012kg}. Furthermore, the recursive techniques developed for supergravity in Refs. \cite{Berends:1988zp,Bern:1998sv}, may be helpful. Nevertheless, it may take quite a tour de force to make a connection between the two sides.

The description \req{smel1} of superstring scattering amplitudes as Mellin transforms of
supergravity amplitudes or generically the inverse Mellin transform \req{FORMB}
of string form factors \req{FORM} into  products of
delta--functions localizing in Mellin position space might point towards
a dual description of perturbative string theory. Basic building blocks of the latter are
graphs and delta--functions assembled by rules coming from the Pascal's triangle in Figure~1.
Moving to Mellin position space, which bypasses Koba--Nielsen factors, might allow  to directly compute on--shell superstring scattering amplitudes without resorting to the conventional evolution through space--time.
From a mathematical point of view the distributional setting of Mellin transformation
converting  string form factors \req{FORM} to Dirac delta--functions  \req{FORMB} in
Mellin position space provides a new ground for studying superstring amplitudes, \eg for recursion relations, by working directly in  position space.

So what if all superstring amplitudes are given by some Mellin transforms of supergravity amplitudes? Can we ``trivialize'' string theory? The key to the answer lies in Figure 1. In order to compute the transform, supergravity amplitudes are uplifted to the Mellin position space and then integrated over a surface constrained by the Pascal's triangle of nonlinear equations. Understanding the nature of this embedding should help unraveling a deeper superstring/supergravity correspondence. Superstring theory may well be supergravity in a brilliant disguise.

%\break

\vskip 3ex\noindent
\textbf{Acknowledgements}\\

St.St. would like to thank the Institute for Advanced Study at Princeton for its hospitality during initiating this work.
This material is based in part upon work supported by the National Science Foundation under Grant No.\ PHY-0757959.  Any
opinions, findings, and conclusions or recommendations expressed in
this material are those of the authors and do not necessarily reflect
the views of the National Science Foundation.

%\break

\appendix
\section{From Cayley graphs to Hamiltonian graphs}

In this Appendix we discuss  polynomial reduction  of the set of rational functions in $z_{ij}$,
which appears in \req{samp1}. The latter can be related to $(N-2)^{(N-4)}$ labelled trees.
We shall prove, that partial fraction decomposition reduces this set of rational functions
to a basis of $(N-3)!$ elements, which appears in Eq. \req{samp2}.

Let us first introduce some common notion in graph theory.
A graph $G$ constitutes a set of vertices $V$ and a set of edges $E$, with each edge $e$
being a pair  of two different vertices $v_1,v_2$ and no more than one edge between two vertices.
A tree graph is a connected graph without cycles. The graph $P_n$ is simply a path on $n$ vertices.
A spanning tree of a connected graph is a tree comprising all vertices.
In a complete graph every two of its vertices are adjacent.
The degree $deg(V)$ of the vertex $V$ is the number of edges attached to it.
A rooted tree is a tree with one vertex designed as a root.
Finally, a Hamiltonian path is a path in an undirected graph, that visits each vertex
exactly once.

The tree diagrams of interest are Cayley graphs $C_n$. The latter describe labelled trees on $n$ vertices. According to Cayley there are $n^{n-2}$ of them \cite{Cayley}.
For $n=3$ we have the three diagrams with vertices $i,j$ and $k$, depicted in Figure \ref{cayley1}:
\begin{figure}[H]
\centering
\includegraphics[width=0.2\textwidth]{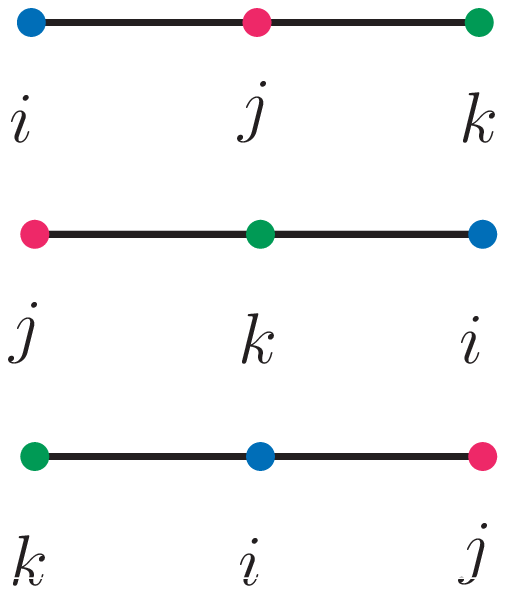}
\caption{Cayley graphs for $n=3$.}
\label{cayley1}
\end{figure}
\noindent
On the other hand, for $n=4$ we have the following sixteen diagrams with vertices $i,j,k$
and~$l$:
\begin{figure}[H]
\centering
\includegraphics[width=0.9\textwidth]{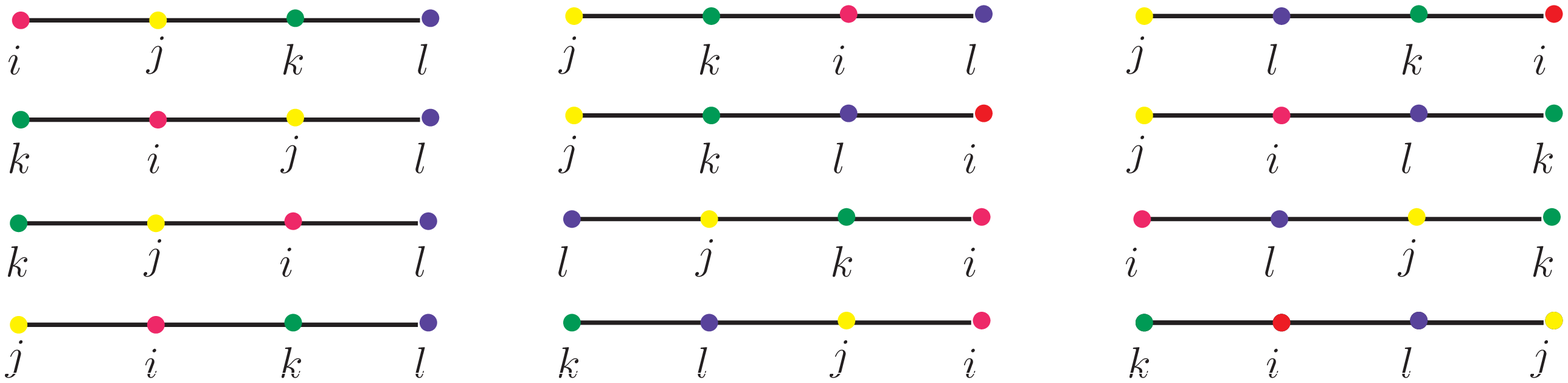}
%\caption{Cayley graphs for $n=4$.}
\label{cayley2a}
\end{figure}
\begin{figure}[H]
\centering
\includegraphics[width=0.9\textwidth]{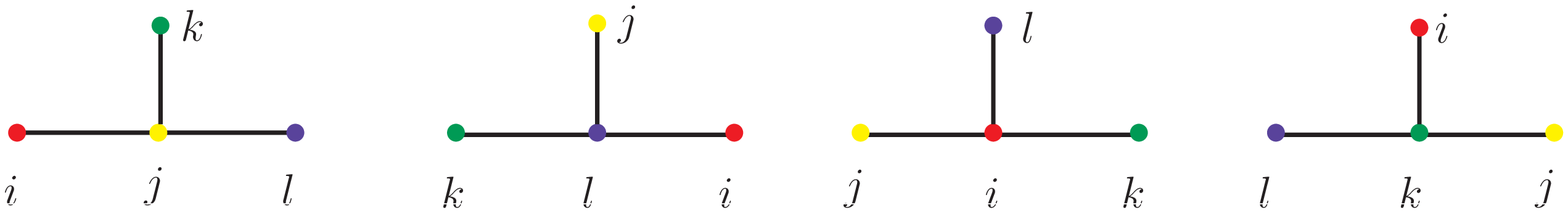}
\caption{Cayley graphs for $n=4$.}
\label{cayley2b}
\end{figure}

As advocated in Section 2 we assign to each product of rational functions a tree graph, \eg
the following rational function are associated with the corresponding diagrams
\be\label{TREED}
\begin{array}{lcl}
\ds{\fc{1}{z_{ij}z_{jk}z_{kl}}=}&&
\begin{picture}(0,0)(0,-3)
\Line(0,0)(60,0)
\Vertex(20,0){1.75}\Vertex(0,0){1.75}\Vertex(40,0){1.75}\Vertex(60,0){1.75}
\Text(0,-5)[t]{$i$}\Text(20,-5)[t]{$j$}\Text(40,-5)[t]{$k$}\Text(60,-5)[t]{$l$}\Text(70,0)[]{$,$}
\end{picture}\\[12mm]
\ds{\fc{1}{z_{ij}z_{jk}z_{jl}}=}&&
\begin{picture}(0,0)(0,-3)
\Line(0,0)(40,0)\Line(20,0)(20,20)
\Vertex(20,0){1.75}\Vertex(0,0){1.75}\Vertex(40,0){1.75}\Vertex(20,20){1.75}
\Text(0,-5)[t]{$i$}\Text(20,-5)[t]{$j$}\Text(40,-5)[t]{$k$}\Text(20,25)[b]{$l$}\Text(50,0)[]{$,$}
\end{picture}
\end{array}
\ee
respectively.
Partial fraction decomposition on rational functions \req{TREED}
gives rise to relations between trees.
In Section 2 we have attributed to each Cayley graph $C_n$ the corresponding rational function $C_\sigma^N$ in the coordinates $z_i$, with $n=N-2$. Many of them can be related by partial fractioning.
In this Appendix we want to find those functions, which can no longer be related
subject to partial fraction relations, \ie they form a basis. As we shall see their corresponding trees represent a special subset of the full set of labelled trees $C_n$.

In fact, partial fractioning allows to reduce any tree  with vertices comprising several
branchings to a tree diagram with vertices to which at most two edges are attached, \ie trees with
vertices $v$ of degree $deg(v)\geq 3$ can be always be brought to a sum of trees with vertices
$v$ of degree $deg(v)\leq 2$.
{\it E.g.} for a rational function corresponding to a
tree with one vertex $j$ of degree $deg(j)=3$ we have the following decomposition
\be\label{rule1}
\fc{1}{z_{ij}z_{jk}z_{jl}}=\begin{cases}
\ds{\fc{1}{z_{ij}}\ \lf(\fc{1}{z_{jl}z_{lk}}+\fc{1}{z_{jk}z_{kl}}\ri)}\ ,&\\[4mm]
\ds{-\lf(\fc{1}{z_{li}z_{ij}}+\fc{1}{z_{il}z_{lj}}\ri)\ \fc{1}{z_{jk}}}\ ,&
\end{cases}
\ee
which graphically can be depicted as shown in Figure \ref{graph1}:
\begin{figure}[H]
\centering
\includegraphics[width=0.8\textwidth]{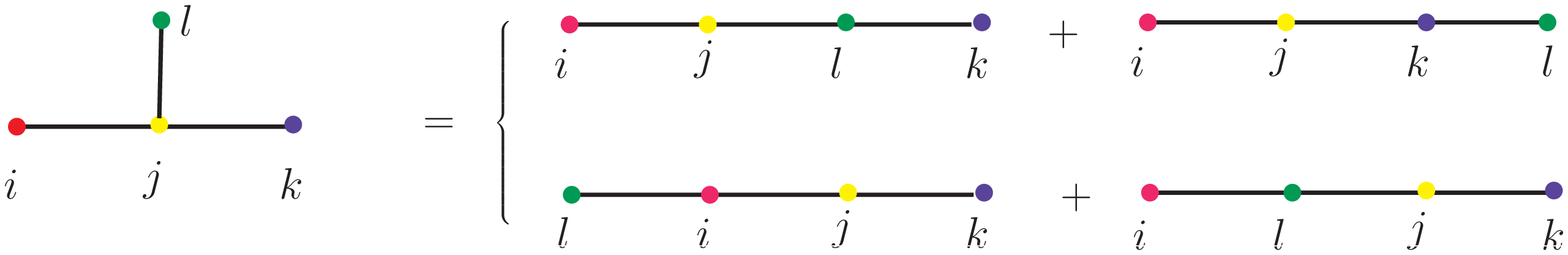}
\caption{Partional fractioning converts a vertex $j$ of  $deg(j)=3$ to a vertex of  $deg(j)=2$.}
\label{graph1}
\end{figure}
\noindent
Hence, we may convert the vertex $j$ of degree $deg(j)=3$ to a vertex of degree $deg(j)=2$ by
moving the vertex $l$ either to the right hand side of $j$ or to its left hand side.
More generally, for a tree diagram with $n$ vertices $i_l,\ l=1,\ldots,n$ and one vertex $i_2$ having $deg(i_2)=3$ we obtain $n-2$ tree diagrams with $deg(i_2)=2$ and with the vertex $i_n$ moved to the right hand side of $i_2$.
Alternatively, we obtain two trees by moving the vertex $i_n$ to the left hand side
of the vertex $i_2$:
\vskip0.25cm
\be\label{rule2}
%%%%%%%%%%%%%%%%%%%%%%%%%%%%%%%%%%%%%%%%%%%%%%%%%%%%%%%%%%%%%%%%%%%
\begin{picture}(0,0)(150,0)
\Line(0,0)(70,0)\Line(20,0)(20,20)\DashLine(70,0)(100,0){1.5}
\Line(100,0)(110,0)
\Vertex(20,0){1.75}\Vertex(0,0){1.75}\Vertex(40,0){1.75}\Vertex(60,0){1.75}
\Vertex(110,0){1.75}\SetColor{Red}\Vertex(20,20){1.75}
\Text(0,-5)[t]{$i_1$}\Text(20,-5)[t]{$i_2$}\Text(40,-5)[t]{$i_3$}\Text(20,25)[b]{$i_n$}\Text(60,-5)[t]{$i_4$}\Text(110,-5)[t]{$i_{n-1}$}
\end{picture}
%%%%%%%%%%%%%%%%%%%%%%%%%%%%%%%%%%%%%%%%%%%%%%%%%%%%%%%%%%%%%%%%%%%
\hskip-1cm=\begin{cases}
\begin{picture}(120,0)(0,-5)
\Text(20,0)[]{$\sum\limits_{l=2}^{n-1}$}
\Line(40,0)(90,0)\DashLine(90,0)(110,0){1.5}
\Line(110,0)(170,0)\DashLine(170,0)(190,0){1.5}\Line(190,0)(220,0)
\Vertex(60,0){1.75}\Vertex(40,0){1.75}\Vertex(80,0){1.75}
\Vertex(120,0){1.75}\Vertex(160,0){1.75}
\Vertex(200,0){1.75}\Vertex(220,0){1.75}
\SetColor{Red}\Vertex(140,0){1.75}\Text(40,-5)[t]{$i_1$}\Text(60,-5)[t]{$i_2$}\Text(80,-5)[t]{$i_3$}\Text(120,-5)[t]{$i_{l}$}
\Text(140,-5)[t]{$i_n$}\Text(160,-5)[t]{$i_{l+1}$}\Text(200,-5)[t]{$i_{n-2}$}\Text(220,-5)[lt]{$i_{n-1}$}\Text(235,0)[t]{$,$}
\end{picture}&\\[5mm]
%%%%%%%%%%%%%%%%%%%%%%%%%%%%%%%%%%%%%%%%%%%%%%%%%%%%%%%%%%%%%%%%%%%
%%%%%%%%%%%%%%%%%%%%%%%%%%%%%%%%%%%%%%%%%%%%%%%%%%%%%%%%%%%%%%%%%%%
\begin{picture}(120,0)(0,0)
\Text(10,0)[]{$-$}
\Line(30,0)(80,0)\DashLine(80,0)(110,0){1.5}\Line(110,0)(120,0)
\Vertex(50,0){1.75}\Vertex(70,0){1.75}\Vertex(120,0){1.75}
\Text(30,-5)[t]{$i_n$}\Text(50,-5)[t]{$i_1$}\Text(70,-5)[t]{$i_2$}\Text(120,-5)[t]{$i_{n-1}$}
\Text(145,0)[]{$-$}
%%%%%%%%%%%%%%%%%%%%%%%%%%%%%%%%%%%%%%%%%%%%%%%%%%%%%%%%%%%%%%%%%%%
\Line(170,0)(220,0)\DashLine(220,0)(250,0){1.5}\Line(250,0)(260,0)
\Vertex(170,0){1.75}\Vertex(210,0){1.75}
\Vertex(260,0){1.75}
\Text(170,-5)[t]{$i_1$}\Text(190,-5)[t]{$i_n$}\Text(210,-5)[t]{$i_2$}\Text(260,-5)[t]{$i_{n-1}$}
\SetColor{Red}\Vertex(30,0){1.75}\Vertex(190,0){1.75}
\end{picture}&\end{cases}\\[5mm]
%%%%%%%%%%%%%%%%%%%%%%%%%%%%%%%%%%%%%%%%%%%%%%%%%%%%%%%%%%%%%%%%%%%
\ee
In terms of partional fraction relations the tree diagrams \req{rule2} describe the two decompositions:
\be
\fc{1}{z_{i_1i_2}z_{i_2i_3}\ldots z_{i_{n-2}i_{n-1}}z_{i_2i_n}}=\begin{cases}
\hskip0.25cm\sum\limits_{l=2}^{n-1}
\fc{1}{z_{i_1i_2}z_{i_2i_3}\ldots z_{i_li_n}z_{i_ni_{l+1}}\ldots  z_{i_{n-2}i_{n-1}}}\ ,&\\[5mm]
-\fc{1}{z_{i_ni_1}z_{i_1i_2}z_{i_2i_3}\ldots z_{i_{n-2}i_{n-1}}}-
\fc{1}{z_{i_1i_n}z_{i_ni_2}z_{i_2i_3}\ldots z_{i_{n-2}i_{n-1}}}\ ,&\end{cases}\\[5mm]
%%%%%%%%%%%%%%%%%%%%%%%%%%%%%%%%%%%%%%%%%%%%%%%%%%%%%%%%%%%%%%%%%%%
\ee
respectively. Three comments need to be made in the following.
The above reasoning has been established for the vertex $i_n$ having degree $deg(i_n)=1$, \ie
no further edges are attached to it.
If the degree of the vertex $i_n$ was $deg(i_n)=d>1$, in most of the diagrams on the right hand side of \req{rule2} the degree of this vertex would become $deg(i_n)=d+1>2$.
However, in this case  by repeated use of  \req{rule2} the multiple branching at $i_n$
can eventually be removed ending up at diagrams with $deg(i_n)\leq 2$ on the right hand side
of \req{rule2}.
Furthermore, if for the vertex $i_2$ the degree was $deg(i_2)=d>3$ by applying \req{rule2}
once we would get $deg(i_2)=d-1$ in the diagrams on the right hand side.
Again, repeated use of  \req{rule2} eventually provides diagrams for which $deg(i_2)=2$.
Finally, if there were branchings not only at the vertex $i_2$ but also at other vertices,  \ie $deg(i_l)>2$, $i_l,\ l=3,\ldots,n-2$, the above reasoning applies as well.

To conclude, by repeated application of  \req{rule2} any rational function associated to one of the $n^{n-2}$ Cayley trees $C_n$ can be  reduced to a sum of rational functions corresponding to
tree diagrams $P_n$ with vertices of degree at most two.
There are $\h n!$ of the latter.
Hence, in the following we only consider those $\h (N-2)!$ rational functions, which are
associated to tree diagrams $P_n$.
{}From those diagrams we can single out rooted trees with the vertex $a$ designated as a root,
\ie $deg(a)=1$.
Let us consider the following rational function
\be\label{consider2}
\fc{1}{z_{ij}z_{ja}z_{ak}}=
\begin{picture}(0,0)(0,-3)
\Line(0,0)(60,0)
\Vertex(20,0){1.75}\Vertex(0,0){1.75}\Vertex(40,0){1.75}\Vertex(60,0){1.75}
\Text(0,-5)[t]{$i$}\Text(20,-5)[t]{$j$}\Text(40,-5)[t]{$a$}\Text(60,-5)[t]{$k$}
\Text(70,0)[]{$,$}
\end{picture}
\ee
which after performing partial fraction decomposition becomes:
\be\label{consider3}
\fc{1}{z_{ij}z_{jk}z_{ak}}+\fc{1}{z_{ij}z_{jk}z_{ja}}=
\begin{picture}(0,0)(0,-3)
\Line(0,0)(60,0)
\Vertex(20,0){1.75}\Vertex(0,0){1.75}\Vertex(40,0){1.75}\Vertex(60,0){1.75}
\Text(0,-5)[t]{$i$}\Text(20,-5)[t]{$j$}\Text(40,-5)[t]{$k$}\Text(60,-5)[t]{$a$}
\Text(70,0)[]{$+$}
\Vertex(80,0){1.75}\Vertex(100,0){1.75}\Vertex(120,0){1.75}\Line(100,0)(100,20)\Line(80,0)(120,0)
\Text(80,-5)[t]{$i$}\Text(100,-5)[t]{$j$}\Text(120,-5)[t]{$a$}\Text(100,25)[b]{$k$}
\Vertex(100,20){1.75}\Text(130,0)[]{$.$}
\end{picture}
\ee
With \req{rule2} we may get rid of the last diagram to arrive at:
\be\label{consider4}
\hskip-9cm\fc{1}{z_{ij}z_{jk}z_{ak}}-\fc{1}{z_{ik}z_{kj}z_{ja}}-\fc{1}{z_{ki}z_{ij}z_{ja}}=
\begin{picture}(0,0)(0,-3)
\Line(0,0)(60,0)
\Vertex(20,0){1.75}\Vertex(0,0){1.75}\Vertex(40,0){1.75}\Vertex(60,0){1.75}
\Text(0,-5)[t]{$i$}\Text(20,-5)[t]{$j$}\Text(40,-5)[t]{$k$}\Text(60,-5)[t]{$a$}
\Text(70,0)[]{$+$}
\Vertex(80,0){1.75}\Vertex(100,0){1.75}\Vertex(120,0){1.75}\Line(80,0)(140,0)\Vertex(140,0){1.75}
\Text(80,-5)[t]{$i$}\Text(100,-5)[t]{$k$}\Text(120,-5)[t]{$j$}\Text(140,-5)[t]{$a$}
\Text(150,0)[]{$+$}
\Vertex(160,0){1.75}\Vertex(180,0){1.75}\Vertex(200,0){1.75}\Line(160,0)(220,0)\Vertex(220,0){1.75}
\Text(160,-5)[t]{$k$}\Text(180,-5)[t]{$i$}\Text(200,-5)[t]{$j$}\Text(220,-5)[t]{$a$}
\Text(230,0)[]{$.$}
\end{picture}
\ee
Eqs. \req{consider2}--\req{consider4} demonstrate, that the rational function corresponding to the diagram \req{consider2} can always be written as a sum of rational functions \req{consider4} referring to rooted trees with vertex $a$ as their root.
Hence, in the  tree diagram \req{consider2}
the vertex $a$ can always be moved to the boundary of the tree diagram.
The above reasoning has been established for a tree $P_n$ with $n=4$.
However, the same arguments apply for any tree diagram $P_n$ with vertex $a$ not at the boundary
of the tree: if in \req{consider2} there was
a path on some vertices attached to the vertex $k$ by \req{rule2} in \req{consider4} those vertices can be moved to the left hand side of the three diagrams. The same argument can be used in the case of a path on some vertices attached to vertex $i$.
Note, that during the step from \req{consider3} to \req{consider4} in the second diagram of \req{consider3} according to \req{rule2} we only could move
the vertex $k$ to the left in order to leave the vertex $a$ at the boundary.
This is way we cannot repeat the steps \req{consider2}--\req{consider4} to single out a second vertex and move it to the boundary.

\begin{figure}[H]
\centering
\includegraphics[width=0.6\textwidth]{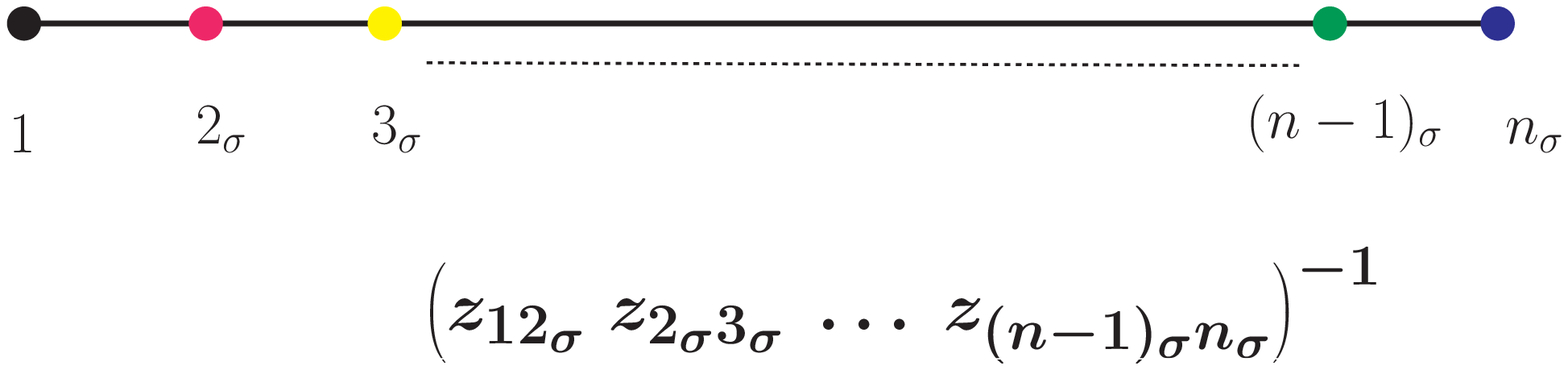}
\caption{Hamilton graphs corresponding to a  minimal basis subject to partial fractionning.}
\label{hamilton}
\end{figure}

To summarize, in the tree diagrams $P_n$ we can single out one vertex $a$
allowing to focus on rooted tree graphs with $a$ designated as $a$ root
connected by a path on the remaining $n-1$ vertices. The latter can be permuted, hence in total there are $(n-1)!$ of them, shown in Figure \ref{hamilton}.
Their corresponding rational functions comprise a minimal basis subject to partial fraction decomposition. Hence, all rational functions described by Cayley graphs can be written in terms
of a $(n-1)!$--dimensional basis corresponding to rooted trees of $P_n$.
\vskip0.45cm

\section{Multiple Mellin transforms for superstrings}

In the superstring amplitude \req{samp1} the sum over $(N-2)^{(N-4)}$
Cayley graphs amounts to considering
for each graph a certain integral over the $N$ world--sheet coordinates $z_i$
\be\label{INTz}
\int d\tilde\mu^S_N(z,s)\ R(\{z_{kl}\})\ \ \ ,\ \ \ R(\{z_{kl}\})=\prod_{\rm edges}
\fc{s_{ij}}{z_{ij}}\ ,
\ee
whose rational function $R(\{z_{kl}\})$ is graphically described by  edges and vertices labelled
by $3,4,\ldots,N$, \cf Appendix A.
To derive \req{smel1} we wrote the integral \req{INTz} as Mellin transform~\req{zmel}
\be\label{INTm}
\int dM_N(u,s)\ M(\{u_{p,q}\})\ \ \ ,\ \ \ M(\{u_{p,q}\})=\prod_{\rm edges}
\sigma_{3j}\ \p\ln\delta(\alpha_{ij})
\ee
of a function $M(\{u_{p,q}\})$ in Mellin position space $u_{i,j}$,
which in turn according to \req{FORMB} gives rise to the correlation function
\be\label{inturn}
G(\{u_{p,q}\})=\lf(\prod_{i,j\in P}\theta(1-u_{i,j})\ri) \
\lf(\prod_{l=4}^{N}\sigma_{3l}^{-1}\ \rho_{l-1,l}\ri)\ \delta(\{u_{p,q}\})\ M(\{u_{p,q}\})
\ee
in inverse Mellin space.
Alternatively,  according to \req{idneti} the form factors  \req{INTz} can be considered
as Mellin amplitudes \req{FORM}, whose inverse Mellin transform \req{FORMB} must agree with \req{inturn}.

For a given form factor $B_N(\{s_{k,l}\},\{n_{k,l}\})$ a momentum insertions $s_{p,q}$
(with $(p,q)\in P$) can easily be incorporated in its inverse Mellin transform \req{FORMB} by acting on the latter
through the respective derivatives in position space $u_{p,q}$ as:
\bea\label{DIFF}
&&(2\pi i)^{-m}\ \lf(\prod_{(i,j)\in P}\  \int_{-i\infty+c}^{+i\infty+c} ds_{i,j} \ u_{i,j}^{-s_{i,j}}\ri)\ s_{p,q}\ B_N(\{s_{k,l}\},\{n_{k,l}\})\nonumber\\[2mm]
&=&-(2\pi i)^{-m}\ u_{p,q}\ \lf\{\fc{\p}{\p u_{p,q}}\ \lf(\prod_{(i,j)\in P}\  \int_{-i\infty+c}^{+i\infty+c} ds_{i,j} \ u_{i,j}^{-s_{i,j}}\ri)\ B_N(\{s_{k,l}\},\{n_{k,l}\})\ri\}\nonumber\\[2mm]
&=&-u_{p,q}\ \fc{\p}{\p u_{p,q}}\ \lf\{\lf(\prod_{(i,j)\in P}\ u_{i,j}^{n_{i,j}}\
\theta(1-u_{i,j})\ri)\ \delta(\{u_{k,l}\})\ri\}\ .
\eea
This way partial integration relations between different form factors
become partial integration relations in Mellin position space.

\tocless\subsection{Mellin transforms for $\bm{N=5}$}

Let us consider the Cayley basis for $N=5$.
In the following we use: $x_1:=u_{1,2},\ x_2:=u_{2,3},\ x_3=u_{3,4},\ x_4=u_{4,5},\ x_5=u_{5,6}$.
The Mellin measure \req{zmel} is:
\be
\int dM_5(u,s)=\lf(\prod_{i=1}^5\int_0^\infty \ x_i^{s_{i,i+1}}\ \theta(1-x_i)\ri)
\fc{x_1}{x_2^3\ x_5^2}\ \delta(\{x_j\})\ ,
\ee
with the corresponding product of delta--functions \req{deltaM}:
\bea\label{deltafactor5}
\delta(\{x_j\})&=&x_1\ \delta(\alpha_{34})\ \delta(\alpha_{45})\ \delta(\alpha_{35})\nonumber\\
&=&x_1\ \delta(1-x_2-x_1x_3)\ \delta(1-x_5-x_1x_4)\ \delta(1-x_1-x_2x_5)\ .
\eea

In the following Table I for each Cayley tree--graph and its corresponding integral  \req{INTz},
we display its Mellin representation \req{INTm} and the underlying correlation function
\req{inturn} in inverse Mellin space.

\begin{table}[H]
\label{tab1}
\centering
\begin{tabular}{|c|c|c|c|c|}
\hline
{\rm \ tree--graph\ \ }
&$R$
&$\ds{\ \prod\limits_{\rm edges}\sigma_{3j}\ }$
&$M$
&{\rm inverse Mellin space} $G$\\
\hline\hline
&&&&\\[-5mm]
%%%%%%%%%%%%%%%%%%%%%%%%%%%%%%
\begin{picture}(40,0)(0,-5)
\Line(0,0)(40,0)
\Vertex(20,0){1.75}\Vertex(0,0){1.75}\Vertex(40,0){1.75}
\Text(0,-5)[t]{${\scriptstyle 3}$}
\Text(20,-5)[t]{$\scriptstyle 4$}
\Text(40,-5)[t]{$\scriptstyle 5$}
\end{picture}
&$\ds{\ \fc{s_{34}s_{45}}{z_{34}z_{45}}\ }$
&$\si_{34}\si_{35}$
&$\ds{\ x_2^2x_5\ \fc{\delta'(\al_{34})}{\delta(\al_{34})} \fc{\delta'(\al_{45})}{\delta(\al_{45})}}$
&$\ds{\ x_1^3x_3x_4\ \delta'(\al_{34})\ \delta'(\al_{45})\ \delta(\al_{35})\  }$\\[5mm]
%%%%%%%%%%%%%%%%%%%%%%%%%%%%%%
\begin{picture}(40,0)(0,-5)
\Line(0,0)(40,0)
\Vertex(20,0){1.75}\Vertex(0,0){1.75}\Vertex(40,0){1.75}
\Text(0,-5)[t]{$\scriptstyle 4$}
\Text(20,-5)[t]{$\scriptstyle 5$}
\Text(40,-5)[t]{$\scriptstyle 3$}
\end{picture}
&$\ds{\ \fc{s_{45}s_{35}}{z_{45}z_{53}}\ }$
&$\si_{35}^2$
&$\ds{\ x_2^2x_5^2\ \fc{\delta'(\al_{45})}{\de(\al_{45})}\
\fc{\de'(\al_{35})}{\de(\al_{35})}\ }$
&$\ds{\ x_1^3x_3x_4x_5\ \delta(\al_{34})\ \delta'(\al_{45})\ \delta'(\al_{35})\ }$\\[5mm]
%%%%%%%%%%%%%%%%%%%%%%%%%%%%%%
\begin{picture}(40,0)(0,-5)
\Line(0,0)(40,0)
\Vertex(20,0){1.75}\Vertex(0,0){1.75}\Vertex(40,0){1.75}
\Text(0,-5)[t]{$\scriptstyle 5$}
\Text(20,-5)[t]{$\scriptstyle 3$}
\Text(40,-5)[t]{$\scriptstyle 4$}
\end{picture}
&$\ds{\ \fc{s_{34}s_{35}}{z_{34}z_{53}}\ }$
&$\si_{34}\si_{35}$
&$\ds{\ x_2^2x_5\ \fc{\delta'(\al_{34})}{\delta(\al_{34})} \fc{\delta'(\al_{35})}{\delta(\al_{35})}\ }$
&$\ds{\ x_1^3x_3x_4\ \delta'(\al_{34})\ \delta(\al_{45})\ \delta'(\al_{35})\ }$\\[5mm]
\hline
\end{tabular}
\caption{Cayley basis and inverse Mellin transform for $N=5$.} %title of the table
\end{table}

Alternatively,  considering  the integrals \req{INTz} as Mellin amplitudes \req{FORM}
gives the following dictionary, depicted in Table II.
\begin{table}[H]
\centering
\begin{tabular}{|c|c|c|c|}
\hline
formfactor
& $z$--space:
& $u$--space:
& inverse Mellin space:\\
\
&$\fc{1}{\prod\limits_{i=1}^5 |z_i-z_{i+2}|}\
\prod\limits_{i<j}^5  z_{ij}^{\widetilde n_{ij}}$ in Eq. \req{koba}
&$\prod\limits_{(i,j)\in P} u_{i,j}^{n_{i,j}-1}$ in Eq. \req{FORM}
&$\prod\limits_{(i,j)\in P} u_{i,j}^{n_{i,j}}$ in Eq. \req{FORMB}\\
\hline\hline
&&&\\[-5mm]
%%%%%%%%%%%%%%%%%%%%%%%%%%%%%%
$\fc{1}{z_{34}z_{45}}$
&$z_{12}^{-1}z_{23}^{-1}z_{34}^{-1}z_{45}^{-1}z_{51}^{-1}$
&$x_1^{-1}x_2^{-1}x_3^{-1}x_4^{-1}x_5^{-1}$
&$1$\\
%%%%%%%%%%%%%%%%%%%%%%%%%%%%%%
$\fc{1}{z_{45}z_{53}}$
&$z_{12}^{-1}z_{23}^{-1}z_{14}^{-1}z_{35}^{-1}z_{45}^{-1}$
&$x_1^{-1}x_2^{-1}x_4^{-1}$
&$x_3\ x_5$\\
%%%%%%%%%%%%%%%%%%%%%%%%%%%%%%
$\fc{1}{z_{34}z_{53}}$
&$z_{13}z_{12}^{-1}z_{23}^{-1}z_{14}^{-1}z_{34}^{-1}z_{15}^{-1}z_{35}^{-1}$
&$x_1^{-1}x_2^{-1}x_3^{-1}x_5^{-1}$
&$x_4$\\
\hline
\end{tabular}
\label{tabii}
\caption{Cayley basis and corresponding Mellin amplitudes for $N=5$.} %title of the table
\end{table}
\noindent
With the information displayed in Table II
we can compute the inverse Mellin transform \req{FORMB} of the formfactors
supplemented by the momentum insertions.
Following the rule \req{DIFF} we obtain from the last column of Table II
\begin{align*}
\ds{\fc{s_{34}s_{45}}{z_{34}z_{45}}}&\simeq\ds{x_3\fc{\p}{\p x_3}\ x_4\fc{\p}{\p x_4}\
\ \lf\{\lf(\prod_{i=1}^5\ \theta(1-x_i)\ri)\ \delta(\{x_j\})\ri\}}\\[3mm]
&=\ds{x_1^3x_3x_4\ \delta'(1-x_2-x_1x_3)\ \delta'(1-x_5-x_1x_4)\ \delta(1-x_1-x_2x_5)\ ,}\\[5mm]
\ds{\fc{s_{35}s_{45}}{z_{35}z_{45}}}&\simeq\ds{x_4\fc{\p}{\p x_4}\
\lf(x_1\fc{\p}{\p x_1}-x_3\fc{\p}{\p x_3}-x_4\fc{\p}{\p x_4}\ri)
\ \lf\{\lf(\prod_{i=1}^5\ \theta(1-x_i)\ri)\ x_3x_5\ \delta(\{x_j\})\ri\}}\\[3mm]
&=\ds{x_1^3x_3x_4x_5\ \delta(1-x_2-x_1x_3)\ \delta'(1-x_5-x_1x_4)\ \delta'(1-x_1-x_2x_5)\ ,}
\\[5mm]
\ds{\fc{s_{34}s_{35}}{z_{34}z_{35}}}&\simeq\ds{x_3\fc{\p}{\p x_3}\
\lf(x_1\fc{\p}{\p x_1}-x_3\fc{\p}{\p x_3}-x_4\fc{\p}{\p x_4}\ri)
\ \lf\{\lf(\prod_{i=1}^5\ \theta(1-x_i)\ri)\ x_4\ \delta(\{x_j\})\ri\}}\\[3mm]
&=\ds{x_1^3x_3x_4\ \delta'(1-x_2-x_1x_3)\ \delta(1-x_5-x_1x_4)\ \delta'(1-x_1-x_2x_5)\ ,}
\end{align*}
in agreement with the last column of Table I.

\tocless\subsection{Mellin transforms for $\bm{N=6}$}

Let us consider the $N=6$ Cayley basis and define:
$x_1:=u_{1,2},\ x_2:=u_{2,3},\ x_3=u_{3,4},\ x_4=u_{4,5},\ x_5=u_{5,6},\ x_6=u_{1,6}$
and $y_1:=u_{1,3},\ y_2:=u_{2,4},\ y_3:=u_{3,5}$.
The Mellin measure \req{zmel} is:
\bea
\int dM_6(u,s)&=&\lf(\prod_{i=1}^6\int_0^\infty \ x_i^{s_{i,i+1}}\ \theta(1-x_i)\ri)\
\lf(\prod_{i=1}^3\int_0^\infty \ y_i^{s_{i,i+2}}\ \theta(1-y_i)\ri)\nonumber \\[3mm]
&\times&\fc{x_1^2\ y_1\ y_3}{x_2^4\ x_6^2\ y_2^3}\ \delta(\{x_i\},\{y_j\})\ ,
\eea
with the corresponding product of delta--functions \req{deltaM}
\bea\label{deltafactor6}
\delta(\{x_i\},\{y_j\})
%% &=& x_1^3y_1y_3\ \de(\al_{35})\ \de(\al_{46})\ \de(\al_{36})\ \de(\al_{56})\
%% \de(\al_{34})\ \de(\al_{45})\nonumber\\
&=&x_1^3y_1y_3\ \de_1\ \de_2\ \de_3\ \de_4\ \de_5\ \de_6\ ,
\eea
with:
\bea
\delta_1&:=&\de(\al_{35})=\delta(x_1y_3+x_2y_2-1)\ \ \ ,\ \ \
\delta_2:=\de(\al_{46})=\delta(x_1y_1+x_6y_2-1)\ ,\nonumber\\
\delta_3&:=&\de(\al_{36})=\delta(x_2x_6y_2+x_1-1)\ \ \ ,\ \ \
\delta_4:=\de(\al_{56})=\delta(x_1x_5y_1+x_6-1)\ ,\nonumber\\
\delta_5&:=&\de(\al_{34})=\delta(x_1x_3y_3+x_2-1)\ \ \ ,\ \ \
\delta_6:=\de(\al_{45})=\delta(x_1x_4y_1y_3+y_2-1)\ .
\eea

In the following Table III for each Hamilton tree--graph and its corresponding integral  \req{INTz},
we display its Mellin representation \req{INTm} and the underlying correlation function
\req{inturn} in inverse Mellin space.

\begin{table}[H]
\label{tab3}
\hskip-0.5cm
%\centering
\begin{tabular}{|c|c|c|c|c|}
\hline
{\rm \ \hskip0.5cm tree--graph\ \hskip0.5cm\ }
&$R$
&$\ds{\ \prod\limits_{\rm edges}\sigma_{3j}\ }$
&$M$
&{\rm inverse Mellin space} $G$\\
\hline\hline
&&&&\\[-5mm]
%%%%%%%%%%%%%%%%%%%%%%%%%%%%%%
\begin{picture}(60,0)(0,-5)
\Line(0,0)(60,0)
\Vertex(20,0){1.75}\Vertex(0,0){1.75}\Vertex(40,0){1.75}\Vertex(60,0){1.75}
\Text(0,-5)[t]{${\scriptstyle 3}$}
\Text(20,-5)[t]{$\scriptstyle 4$}
\Text(40,-5)[t]{$\scriptstyle 5$}
\Text(60,-5)[t]{$\scriptstyle 6$}
\end{picture}
&$\ds{\ \fc{s_{34}s_{45}s_{56}}{z_{34}z_{45}z_{56}}\ }$
&$\si_{34}\si_{35}\si_{36}$
&$\ds{\ x_2^3x_6y_2^2\ \fc{\delta'(\al_{34})}{\delta(\al_{34})} \fc{\delta'(\al_{45})}{\delta(\al_{45})}}\fc{\delta'(\al_{56})}{\delta(\al_{56})}$
&$\ds{\ x_1^6x_3x_4x_5y_1^3y_3^3\
\delta_1\delta_2\delta_3\delta_4'\delta_5'\delta_6'\  }$\\[5mm]
%%%%%%%%%%%%%%%%%%%%%%%%%%%%%%
\begin{picture}(60,0)(0,-5)
\Line(0,0)(60,0)
\Vertex(20,0){1.75}\Vertex(0,0){1.75}\Vertex(40,0){1.75}\Vertex(60,0){1.75}
\Text(0,-5)[t]{${\scriptstyle 3}$}
\Text(20,-5)[t]{$\scriptstyle 4$}
\Text(40,-5)[t]{$\scriptstyle 6$}
\Text(60,-5)[t]{$\scriptstyle 5$}
\end{picture}
&$\ds{\ \fc{s_{34}s_{46}s_{56}}{z_{34}z_{46}z_{56}}\ }$
&$\si_{34}\si_{36}^2$
&$\ds{\ x_2^3x_6^2y_2^2\ \fc{\delta'(\al_{34})}{\de(\al_{34})}\
\fc{\de'(\al_{46})}{\de(\al_{46})}\fc{\delta'(\al_{56})}{\de(\al_{56})}\ }$
&$\ds{\ x_1^6x_3x_4x_5x_6y_1^3y_3^3\
\delta_1\delta_2'\delta_3\delta_4'\delta_5'\delta_6\ }$\\[5mm]
%%%%%%%%%%%%%%%%%%%%%%%%%%%%%%
\begin{picture}(60,0)(0,-5)
\Line(0,0)(60,0)
\Vertex(20,0){1.75}\Vertex(0,0){1.75}\Vertex(40,0){1.75}\Vertex(60,0){1.75}
\Text(0,-5)[t]{${\scriptstyle 3}$}
\Text(20,-5)[t]{$\scriptstyle 6$}
\Text(40,-5)[t]{$\scriptstyle 5$}
\Text(60,-5)[t]{$\scriptstyle 4$}
\end{picture}
&$\ds{\ \fc{s_{36}s_{45}s_{56}}{z_{36}z_{45}z_{56}}\ }$
&$\si_{35}\si_{36}^2$
&$\ds{\ x_2^3x_6^2y_2^3\ \fc{\delta'(\al_{36})}{\delta(\al_{36})} \fc{\delta'(\al_{56})}{\delta(\al_{56})}\fc{\delta'(\al_{45})}{\delta(\al_{45})}\ }$
&$\ds{\ x_1^6x_3x_4x_5x_6y_1^3y_2y_3^3\
\delta_1\delta_2\delta_3'\delta_4'\delta_5\delta_6'\ }$\\[5mm]
%%%%%%%%%%%%%%%%%%%%%%%%%%%%%%
\begin{picture}(60,0)(0,-5)
\Line(0,0)(60,0)
\Vertex(20,0){1.75}\Vertex(0,0){1.75}\Vertex(40,0){1.75}\Vertex(60,0){1.75}
\Text(0,-5)[t]{${\scriptstyle 3}$}
\Text(20,-5)[t]{$\scriptstyle 5$}
\Text(40,-5)[t]{$\scriptstyle 4$}
\Text(60,-5)[t]{$\scriptstyle 6$}
\end{picture}
&$\ds{\ \fc{s_{35}s_{45}s_{46}}{z_{35}z_{45}z_{46}}\ }$
&$\si_{35}^2\si_{36}$
&$\ds{\ x_2^3x_6y_2^3\ \fc{\delta'(\al_{35})}{\delta(\al_{35})} \fc{\delta'(\al_{45})}{\delta(\al_{45})}}\fc{\delta'(\al_{46})}{\delta(\al_{46})}$
&$\ds{\ x_1^6x_3x_4x_5y_1^3y_2y_3^3\
\delta_1'\delta_2'\delta_3\delta_4\delta_5\delta_6'\ }$\\[5mm]
%%%%%%%%%%%%%%%%%%%%%%%%%%%%%%
\begin{picture}(60,0)(0,-5)
\Line(0,0)(60,0)
\Vertex(20,0){1.75}\Vertex(0,0){1.75}\Vertex(40,0){1.75}\Vertex(60,0){1.75}
\Text(0,-5)[t]{${\scriptstyle 3}$}
\Text(20,-5)[t]{$\scriptstyle 6$}
\Text(40,-5)[t]{$\scriptstyle 4$}
\Text(60,-5)[t]{$\scriptstyle 5$}
\end{picture}
&$\ds{\ \fc{s_{36}s_{45}s_{46}}{z_{36}z_{45}z_{46}}\ }$
&$\si_{35}\si_{36}^2$
&$\ds{\ x_2^3x_6^2y_2^3\ \fc{\delta'(\al_{36})}{\de(\al_{36})}\
\fc{\de'(\al_{46})}{\de(\al_{46})}\fc{\de'(\al_{45})}{\de(\al_{45})}\ }$
&$\ds{\ x_1^6x_3x_4x_5x_6y_1^3y_2y_3^3\
\delta_1\delta_2'\delta_3'\delta_4\delta_5\delta_6'\ }$\\[5mm]
%%%%%%%%%%%%%%%%%%%%%%%%%%%%%%
\begin{picture}(60,0)(0,-5)
\Line(0,0)(60,0)
\Vertex(20,0){1.75}\Vertex(0,0){1.75}\Vertex(40,0){1.75}\Vertex(60,0){1.75}
\Text(0,-5)[t]{${\scriptstyle 3}$}
\Text(20,-5)[t]{$\scriptstyle 5$}
\Text(40,-5)[t]{$\scriptstyle 6$}
\Text(60,-5)[t]{$\scriptstyle 4$}
\end{picture}
&$\ds{\ \fc{s_{35}s_{46}s_{56}}{z_{35}z_{46}z_{56}}\ }$
&$\si_{35}\si_{36}^2$
&$\ds{\ x_2^3x_6^2y_2^3\ \fc{\delta'(\al_{35})}{\de(\al_{35})}\
\fc{\de'(\al_{56})}{\de(\al_{56})}\fc{\de'(\al_{46})}{\de(\al_{46})}}$
&$\ds{\ x_1^6x_3x_4x_5x_6y_1^3y_2y_3^3\
\ \delta_1'\delta_2'\delta_3\delta_4'\delta_5\delta_6 }$\\[5mm]
\hline
\end{tabular}
\caption{Hamilton basis and inverse Mellin transform for $N=6$.} %title of the table
\end{table}

Alternatively,  considering  the integrals \req{INTz} as Mellin amplitudes \req{FORM}
gives the following dictionary, depicted in Table IV.
\begin{table}[H]
\centering
\begin{tabular}{|c|c|c|c|}
\hline
formfactor
& $z$--space:
& $u$--space:
& inverse Mellin space:\\
\
&$\fc{1}{\prod\limits_{i=1}^6 |z_i-z_{i+2}|}\
\prod\limits_{i<j}^6  z_{ij}^{\widetilde n_{ij}}$ in Eq. \req{koba}
&$\prod\limits_{(i,j)\in P} u_{i,j}^{n_{i,j}-1}$ in Eq. \req{FORM}
&$\prod\limits_{(i,j)\in P} u_{i,j}^{n_{i,j}}$ in Eq. \req{FORMB}\\
\hline\hline
&&&\\[-5mm]
%%%%%%%%%%%%%%%%%%%%%%%%%%%%%%
$\fc{1}{z_{34}z_{45}z_{56}}$
&$z_{12}^{-1}z_{23}^{-1}z_{34}^{-1}z_{45}^{-1}z_{56}^{-1}z_{16}^{-1}$
&$x_1^{-1}x_2^{-1}x_3^{-1}x_4^{-1}x_5^{-1}x_6^{-1}y_1^{-1}y_2^{-1}y_3^{-1}$
&$1$\\
%%%%%%%%%%%%%%%%%%%%%%%%%%%%%%
$\fc{1}{z_{34}z_{46}z_{56}}$
&$z_{12}^{-1}z_{23}^{-1}z_{34}^{-1}z_{46}^{-1}z_{56}^{-1}z_{15}^{-1}$
&$x_1^{-1}x_2^{-1}x_3^{-1}x_5^{-1}y_1^{-1}y_2^{-1}$
&$x_4x_6y_3$\\
%%%%%%%%%%%%%%%%%%%%%%%%%%%%%%
$\fc{1}{z_{36}z_{45}z_{56}}$
&$z_{12}^{-1}z_{23}^{-1}z_{36}^{-1}z_{45}^{-1}z_{56}^{-1}z_{14}^{-1}$
&$x_1^{-1}x_2^{-1}x_4^{-1}x_5^{-1}y_1^{-1}$
&$x_3x_6y_2y_3$\\
%%%%%%%%%%%%%%%%%%%%%%%%%%%%%%
$\fc{1}{z_{35}z_{45}z_{46}}$
&$z_{12}^{-1}z_{23}^{-1}z_{35}^{-1}z_{45}^{-1}z_{46}^{-1}z_{16}^{-1}$
&$x_1^{-1}x_2^{-1}x_4^{-1}x_6^{-1}y_1^{-1}y_3^{-1}$
&$x_3x_5y_2$\\
%%%%%%%%%%%%%%%%%%%%%%%%%%%%%%
$\fc{1}{z_{36}z_{45}z_{46}}$
&$z_{12}^{-1}z_{23}^{-1}z_{36}^{-1}z_{45}^{-1}z_{46}^{-1}z_{15}^{-1}$
&$x_1^{-1}x_2^{-1}x_4^{-1}y_1^{-1}$
&$x_3x_5x_6y_2y_3$\\
%%%%%%%%%%%%%%%%%%%%%%%%%%%%%%
$\fc{1}{z_{35}z_{46}z_{56}}$
&$z_{12}^{-1}z_{23}^{-1}z_{35}^{-1}z_{46}^{-1}z_{56}^{-1}z_{14}^{-1}$
&$x_1^{-1}x_2^{-1}x_5^{-1}y_1^{-1}$
&$x_3x_4x_6y_2y_3$\\
\hline
\end{tabular}
\label{tabiv}
\caption{Hamilton basis and corresponding Mellin amplitudes for $N=6$.} %title of the table
\end{table}
\noindent
With the information displayed in Table IV
we can compute the inverse Mellin transform \req{FORMB} of the formfactors
supplemented by the momentum insertions.
Following the rule \req{DIFF} we obtain from the last column of Table IV
\begin{align*}
\ds{\fc{s_{34}s_{45}s_{56}}{z_{34}z_{45}z_{56}}}&\simeq\ds{x_3x_4x_5\fc{\p}{\p x_3}\fc{\p}{\p x_4}\ \fc{\p}{\p x_5}\ \lf\{\lf(\prod_{i=1}^6\ \theta(1-x_i)\ri)\
\lf(\prod_{i=1}^3\ \theta(1-y_i)\ri)\ \delta(\{x_i\})\ \delta(\{y_j\})\ri\}}\\[3mm]
&=\ds{x_1^6x_3x_4x_5y_1^3y_3^3\
\delta_1\delta_2\delta_3\delta_4'\delta_5'\delta_6'\ ,}\\[3mm]
%%%%%%%%%%%%%%%%%%%%%%%%%%%%%%%%%%%%%%%%%%%%%%%%%%%%%%%%%%%%%%%
\ds{\fc{s_{34}s_{46}s_{56}}{z_{34}z_{46}z_{56}}}&\simeq
\ds{x_4x_5\ \fc{\p}{\p x_4}\fc{\p}{\p x_5}\
\lf(y_1\fc{\p}{\p y_1}-x_4\fc{\p}{\p x_4}-x_5\fc{\p}{\p x_5}\ri)}\\[3mm]
&\ds{\ \lf\{\lf(\prod_{i=1}^6\ \theta(1-x_i)\ri)\ \lf(\prod_{i=1}^3\ \theta(1-y_i)\ri)\
x_4x_6y_3\ \delta(\{x_i\})\ \delta(\{y_j\})\ri\}}\\[3mm]
&=\ds{x_1^6x_3x_4x_5x_6y_1^3y_3^3\
\delta_1\delta_2'\delta_3\delta_4'\delta_5'\delta_6\ ,}\\[3mm]
%%%%%%%%%%%%%%%%%%%%%%%%%%%%%%%%%%%%%%%%%%%%%%%%%%%%%%%%%%%%%%%
\ds{\fc{s_{36}s_{45}s_{56}}{z_{36}z_{45}z_{56}}}&\simeq\ds{x_4x_5\
\fc{\p}{\p x_4}\fc{\p}{\p x_5}\ \lf(x_1\fc{\p}{\p x_1}+x_4\fc{\p}{\p x_4}-y_1\fc{\p}{\p y_1}-y_3\fc{\p}{\p y_3}\ri)}\\[3mm]
&\ds{\ \lf\{\lf(\prod_{i=1}^6\ \theta(1-x_i)\ri)\ \lf(\prod_{i=1}^3\ \theta(1-y_i)\ri)\ x_3x_6y_2y_3\ \delta(\{x_i\})\ \delta(\{y_j\})\ri\}}\\[3mm]
&=\ds{x_1^6x_3x_4x_5x_6y_1^3y_2y_3^3\
\delta_1\delta_2\delta_3'\delta_4'\delta_5\delta_6'\ ,}\\[3mm]
%%%%%%%%%%%%%%%%%%%%%%%%%%%%%%%%%%%%%%%%%%%%%%%%%%%%%%%%%%%%%%%%%
\ds{\fc{s_{35}s_{45}s_{46}}{z_{35}z_{45}z_{46}}}&\simeq\ds{x_4\fc{\p}{\p x_4}\
\lf(y_3\fc{\p}{\p y_3}-x_3\fc{\p}{\p x_3}-x_4\fc{\p}{\p x_4}\ri)\
\lf(y_1\fc{\p}{\p y_1}-x_4\fc{\p}{\p x_4}-x_5\fc{\p}{\p x_5}\ri)}\\[3mm]
&\ds{\ \lf\{\lf(\prod_{i=1}^6\ \theta(1-x_i)\ri)\ \lf(\prod_{i=1}^3\ \theta(1-y_i)\ri)\
\delta(\{x_i\})\ \delta(\{y_j\})\ x_3x_5y_2\ri\}}\\[3mm]
&=\ds{x_1^6x_3x_4x_5y_1^3y_2y_3^3\
\delta_1'\delta_2'\delta_3\delta_4\delta_5\delta_6'\ ,}\\[3mm]
%%%%%%%%%%%%%%%%%%%%%%%%%%%%%%%%%%%%%%%%%%%%%%%%%%%%%%%%%%%%%%%
\ds{\fc{s_{36}s_{45}s_{46}}{z_{36}z_{45}z_{46}}}&\simeq\ds{x_4\fc{\p}{\p x_4}
\ \lf(x_1\fc{\p}{\p x_1}+x_4\fc{\p}{\p x_4}-y_1\fc{\p}{\p y_1}-y_3\fc{\p}{\p y_3}\ri)\
\lf(y_1\fc{\p}{\p y_1}-x_4\fc{\p}{\p x_4}-x_5\fc{\p}{\p x_5}\ri)}\\[3mm]
&\ds{\ \lf\{\lf(\prod_{i=1}^6\ \theta(1-x_i)\ri)\ \lf(\prod_{i=1}^3\ \theta(1-y_i)\ri)\ x_3x_5x_6y_2y_3\ \delta(\{x_i\})\ \delta(\{y_j\})\ri\}}\\[3mm]
&=\ds{x_1^6x_3x_4x_5x_6y_1^3y_2y_3^3\
\delta_1\delta_2'\delta_3'\delta_4\delta_5\delta_6'\ ,}\\[3mm]
%%%%%%%%%%%%%%%%%%%%%%%%%%%%%%%%%%%%%%%%%%%%%%%%%%%%%%%%%%%%%%%
\ds{\fc{s_{35}s_{46}s_{56}}{z_{35}z_{46}z_{56}}}&\simeq\ds{x_5\fc{\p}{\p x_5}\
\lf(y_3\fc{\p}{\p y_3}-x_3\fc{\p}{\p x_3}-x_4\fc{\p}{\p x_4}\ri)\
\lf(y_1\fc{\p}{\p y_1}-x_4\fc{\p}{\p x_4}-x_5\fc{\p}{\p x_5}\ri)}\\[3mm]
&\ds{\ \lf\{\lf(\prod_{i=1}^6\ \theta(1-x_i)\ri)\ \lf(\prod_{i=1}^3\ \theta(1-y_i)\ri)\ x_3x_4x_6y_2y_3\ \delta(\{x_i\})\ \delta(\{y_j\})\ri\}}\\[3mm]
&=\ds{x_1^6x_3x_4x_5x_6y_1^3y_2y_3^3\
\delta_1'\delta_2'\delta_3\delta_4'\delta_5\delta_6\ ,}
\end{align*}
in agreement with the last column of Table III.

\section[Inverse Mellin--Barnes transformation for $N=5$]{Inverse Mellin--Barnes transformation for $N=5$}

In this Appendix we prove the equality \req{QUINT}:
\ba
&&\fc{1}{(2\pi i)^5}\
\int\limits_{-i\infty+c}^{+i\infty+c}ds_1\int\limits_{-i\infty+c}^{+i\infty+c}ds_2
\int\limits_{-i\infty+c}^{+i\infty+c}ds_3\int\limits_{-i\infty+c}^{+i\infty+c}ds_4
\int\limits_{-i\infty+c}^{+i\infty+c}ds_5\
u_{1,2}^{-s_1}\ u_{2,3}^{-s_2}\ u_{3,4}^{-s_3}\ u_{4,5}^{-s_4}\ u_{1,5}^{-s_5}\nnn
&&\hskip1cm\times\fc{\Gamma(s_2+n_{23})\ \Gamma(s_3+n_{34})}{\Gamma(s_2+s_3+n_{23}+n_{34})}\
\fc{\Gamma(s_4+n_{45})\ \Gamma(s_5+n_{15})}{\Gamma(s_4+s_5+n_{15}+n_{45})}\nnn
&&\hskip1cm\times\FF{3}{2}\lf[{s_2+n_{23},\ s_5+n_{15},\ s_3+s_4-s_1+n_{34}+n_{45}-n_{12}\atop
s_2+s_3+n_{23}+n_{34},\ s_4+s_5+n_{15}+n_{45}};1\ri]\label{QUINTT}\\[5mm]
&&\hskip1cm=u_{1,2}^{n_{12}}\ u_{2,3}^{n_{23}}\ u_{3,4}^{n_{34}}\ u_{4,5}^{n_{13}}
\ u_{1,5}^{n_{24}}\ \ \delta(u_{4,5}+u_{3,4}u_{1,5}-1)\ \delta(u_{1,2}+u_{2,3}u_{1,5}-1)\nnn
&&\hskip1cm\times \delta(u_{3,4}+u_{2,3}u_{4,5}-1)\ \theta(1-u_{1,2})\ \theta(1-u_{2,3})\ \ \theta(1-u_{3,4})\ \theta(1-u_{4,5})\ \theta(1-u_{1,5})\ .\nonumber
\ea
After writing the hypergeometric function $\FF{3}{2}$ as power series
$$\begin{array}{lcl}
&&\ds{\fc{\Gamma(s_2+n_{23})\ \Gamma(s_3+n_{34})}{\Gamma(s_2+s_3+n_{23}+n_{34})}\
\fc{\Gamma(s_4+n_{45})\ \Gamma(s_5+n_{15})}{\Gamma(s_4+s_5+n_{45}+n_{15})}}\nnn
&\times&\ds{\FF{3}{2}\lf[{s_2+n_{23},\ s_5+n_{15},\ s_3+s_4-s_1+n_{34}+n_{45}-n_{12}\atop
s_2+s_3+n_{23}+n_{34},\ s_4+s_5+n_{15}+n_{45}};1\ri]}\nnn
&=&\ds{\Gamma(s_3+n_{34})\ \Gamma(s_4+n_{45})\
\sum\limits_{n=0}^\infty\fc{1}{n!}\ \fc{\Gamma(s_2+n_{23}+n)}{\Gamma(s_2+s_3+n_{23}+n_{34}+n)}}\\[5mm]
&\times&\ds{\fc{\Gamma(s_5+n_{15}+n)}{\Gamma(s_4+s_5+n_{45}+n_{15}+n)}\
\fc{\Gamma(s_3+s_4-s_1+n_{34}+n_{45}-n_{12}+n)}{\Gamma(s_3+s_4-s_1+n_{34}+n_{45}-n_{12})}\ ,}
\end{array}
$$
in \req{QUINTT} we perform each of the five integrations, separately. The integrations over $s_5,s_2$ and
$s_1$ give
\begin{align*}
\ds{\fc{1}{2\pi i}\  \int_{-i\infty+c}^{+i\infty+c}ds_5}&\ds{\  u_{1,5}^{-s_5}\
\fc{\Gamma(s_5+n_{15}+n)}{\Gamma(s_4+s_5+n_{45}+n_{15}+n)}}\\[5mm]
&=\ds{u_{1,5}^{n_{15}+n}\ \Gamma(s_4+n_{45})^{-1}\ (1-u_{1,5})^{s_4+n_{45}-1}\ \ \ ,\ \ \ 0<u_{1,5}\leq 1\ ,}\\[5mm]
\ds{\fc{1}{2\pi i}\  \int_{-i\infty+c}^{+i\infty+c}ds_2}&\ds{\   u_{2,3}^{-s_2}\
\fc{\Gamma(s_2+n_{23}+n)}{\Gamma(s_2+s_3+n_{23}+n_{34}+n)}}\\[5mm]
&=\ds{u_{2,3}^{n_{23}+n}\ \Gamma(s_3+n_{34})^{-1}\ (1-u_{2,3})^{s_3+n_{34}-1}\ \ \ ,\ \ \ 0<u_{2,3}\leq 1\ ,}\\[5mm]
\ds{\fc{1}{2\pi i}\  \int_{-i\infty+c}^{+i\infty+c}ds_1}&\ds{\  u_{1,2}^{-s_1}\
\fc{\Gamma(\alpha-s_1+n)}{\Gamma(\alpha-s_1)}=u_{1,2}^{1-\alpha}\ \delta^{(n)}(u_{1,2}-1)}\ ,
\\[5mm]
&\alpha=s_3+s_4+n_{34}+n_{45}-n_{12}\ \ \ ,\ \ \ 0<u_{1,2}\leq 1\ ,
\end{align*}
respectively. For the first two integrals we have applied the relation \req{KAMI}, while
for the last integral we have used
\be\label{WHSI}
\fc{1}{2\pi i}\  \int_{-i\infty+c}^{+i\infty+c}ds\ x^{-s}\ a^{s-1-n}\
\fc{\Gamma(s)}{\Gamma(s-n)}=(-1)^n\ \delta^{(n)}(x-a)\ ,\ a>0\ ,
\ee
which follows from the Mellin transformation ($n\geq0$):
\be\label{WHSIL}
\int_0^\infty dx\ x^{s-1}\ \delta^{(n)}(x-a)=
\begin{cases}
(-1)^n\ \fc{\Gamma(s)}{\Gamma(s-n)}\ a^{s-1-n}\ , &   a>0\ ,\\
0\ ,&a\leq 0\ .
\end{cases},
\ee
The above relation \req{WHSIL} can be proven by applying the fundamental equation, which defines derivatives of the delta--function $\delta$
$$\int f(x)\ \delta^{(n)}(x-a)\ dx=-\int\ \fc{\p f}{\p x}\ \delta^{(n-1)}(x-a)\ dx\equiv (-1)^n\ f^{(n)}(a)$$
for any function $f$ which has continuous derivatives at least up to the $n$--th order in some neighbourhood of the point $x=a$ \cite{Hoskins}.

After collecting all $s_3$-- and $s_4$--dependent terms we are left with the following two integrations
$$\begin{array}{lcl}
\ds{\fc{1}{2\pi i}\  \int_{-i\infty+c}^{+i\infty+c}ds_3\ (u_{1,2}u_{3,4})^{-s_3}\
(1-u_{2,3})^{s_3-1}}&=&\ds{\delta(1-u_{2,3}-u_{1,2}u_{3,4})\ \ \ ,\ \ \ u_{1,2}u_{3,4}>0\ ,}\\[5mm]
\ds{\fc{1}{2\pi i}\  \int_{-i\infty+c}^{+i\infty+c}ds_4\ (u_{1,2}u_{4,5})^{-s_4}\
(1-u_{1,5})^{s_4-1}}&=&\ds{\delta(1-u_{1,5}-u_{1,2}u_{4,5})\ \ \ ,\ \ \ u_{1,2}u_{4,5}>0\ ,}

\end{array}$$
respectively. The latter are evaluated by using \req{KAMII}.

Putting together all $n$--independent terms  gives:
\be\begin{array}{lcl}
&&\hskip-0.5cm
\ds{u_{2,3}^{n_{23}}\ u_{1,5}^{n_{15}}\ (1-u_{2,3})^{n_{34}}\ (1-u_{1,5})^{n_{45}}\ u_{1,2}^{1+n_{12}-n_{34}-n_{45}}
\ \delta(1-u_{1,5}-u_{1,2}u_{4,5})\ \delta(1-u_{2,3}-u_{1,2}u_{3,4})}\\
&&\ds{=u_{1,2}^{1+n_{12}}\ u_{2,3}^{n_{23}}\ u_{3,4}^{n_{34}}\ u_{4,5}^{n_{45}}\ u_{1,5}^{n_{15}}\
\delta(1-u_{1,5}-u_{1,2}u_{4,5})\ \delta(1-u_{2,3}-u_{1,2}u_{3,4})\ .}
\end{array}
\label{PUT1}
\ee
The remaining $n$--dependent terms conspire into the sum:
\be
\sum_{n=0}^\infty\fc{1}{n!}\ \delta^{(n)}(u_{1,2}-1)\ (u_{2,3}\ u_{1,5})^n=\delta(u_{1,2}-1+
u_{2,3}u_{1,5})\ .
\label{PUT2}
\ee
For the above sum we have used the relation
\be
\sum_{n=0}^\infty\fc{x^n}{n!}\ \delta^{(n)}(y)=\delta(x+y)\ ,
\ee
which can be derived by first writing $\delta^{(n)}(y)=\fc{\p^n}{\p y^n}\int\limits^{+\infty}_{-\infty}dk\ e^{2\pi i k y}=\int\limits^{+\infty}_{-\infty}dk\ (2\pi ik)^n\ e^{2\pi i k y}$, then evaluate the sum
$\sum\limits_{n=0}^\infty\fc{(2\pi ikx)^n}{n!}=e^{2\pi i kx}$ and eventually perform the integration $\int\limits^{+\infty}_{-\infty}dk\ e^{2\pi i k y}\ e^{2\pi i kx}
=\delta(x+y)$.

Finally, putting \req{PUT1} and \req{PUT2} together gives the final result \req{QUINTT}:
\be\begin{array}{lcl}
&&\ds{u_{1,2}^{1+n_{12}}\ u_{2,3}^{n_{23}}\ u_{3,4}^{n_{34}}\ u_{4,5}^{n_{45}}\
u_{1,5}^{n_{15}}}\\
&\times&\ds{\delta(1-u_{1,5}-u_{1,2}u_{4,5})\ \delta(1-u_{2,3}-u_{1,2}u_{3,4})\ \delta(u_{1,2}-1+u_{2,3}u_{1,5})}\\[3mm]
&=&\ds{u_{1,2}^{n_{12}}\ u_{2,3}^{n_{23}}\ u_{3,4}^{n_{34}}\ u_{4,5}^{n_{45}}\
u_{1,5}^{n_{15}}}\\
&\times&\delta(u_{1,2}+u_{2,3}u_{1,5}-1)\ \delta(u_{3,4}+u_{2,3}u_{4,5}-1)\
\delta(u_{4,5}+u_{3,4}u_{1,5}-1)\ ,
\end{array}
\label{PUTT}
\ee
with the constraints $0<u_{1,2},u_{2,3},u_{3,4},u_{4,5},u_{1,5}\leq 1$.
In \req{PUTT} the last equality follows from  the following $\delta$--function identity:
\bea\label{BOS}
\delta(\{u_{p,q}\})&=&u_{1,2}^{-1}\ \delta\lf(u_{4,5}-\fc{1-u_{1,5}}{1-u_{2,3}u_{1,5}}\ri)\
\delta\lf(u_{3,4}-\fc{1-u_{2,3}}{1-u_{2,3}u_{1,5}}\ri)\
\delta(u_{1,2}-1+u_{2,3}u_{1,5})\nnn
&=&u_{1,2}\ \delta(1-u_{1,5}-u_{1,2}u_{4,5})\ \delta(1-u_{2,3}-u_{1,2}u_{3,4})\
\delta(1-u_{1,2}-u_{2,3}u_{1,5})\nnn
&=&\delta(u_{1,2}+u_{2,3}u_{1,5}-1)\ \delta(u_{3,4}+u_{2,3}u_{4,5}-1)\
\delta(u_{4,5}+u_{3,4}u_{1,5}-1)\ .
\eea
Note, that the last line corresponds to \req{olddelta} for $N=5$, while the second last
originates from \req{deltaM}.

\end{document}